\documentclass[preprint,a4paper,fleqn]{cas-sc}
\usepackage[authoryear]{natbib}
\usepackage[nolist]{acronym}
\usepackage{subfig}
\usepackage{tabularx}
\usepackage{float}
\usepackage[justification=centering]{caption}
\usepackage{hyperref}

\def\tsc#1{\csdef{#1}{\textsc{\lowercase{#1}}\xspace}}
\tsc{WGM}
\tsc{QE}
\tsc{EP}
\tsc{PMS}
\tsc{BEC}
\tsc{DE}

\newcommand{\fix}[1]{\textcolor{black}{ #1}}

\begin{acronym}[]
	\acro{AI}{Artificial Intelligence}
	\acro{ACC}{Adaptive Cruise Control}
	\acro{ADAS}{Advanced Driver Assistance System}
	\acro{CAN}{Controller Area Network}
	\acro{ECU}{Electronic Control Unit}
	\acro{FNN}{Feedforward Neural Network}
	\acro{GDPR}{General Data Protection Regulation}
	\acro{GOMS}{Goals, Operators, Methods, Selection rules}
	\acro{HCI}{Human-Computer Interaction}
	\acro{HMI}{Human-Machine Interface}
	\acro{HU}{Head Unit}
	\acro{IVIS}{In-Vehicle Information System}
	\acro{KLM}{Keystroke-Level Model}
	\acro{KPI}{Key Performance Indicator}
	\acro{MGD}{Mean Glance Duration}
	\acro{OEM}{Original Equipment Manufacturer}
	\acro{OTA}{Over-The-Air}
	\acro{AOI}{Area of Interest}
	\acro{RF}{Random Forest}
	\acro{SA}{Steering Assist}
	\acro{SHAP}{SHapley Additive exPlanation}
	\acro{TGD}{Total Glance Duration}
	\acro{UCD}{User-centered Design}
	\acro{UX}{User Experience}
	\acro{XAI}{Explainable AI}
\end{acronym}

\begin{document}
\let\WriteBookmarks\relax
\def\floatpagepagefraction{1}
\def\textpagefraction{.001}

\shorttitle{Explainable Predictions for the Visual Demand of In-Vehicle Touchscreen Interactions}

\shortauthors{Ebel et~al.}

\title [mode = title]{On the Forces of Driver Distraction: Explainable Predictions for the Visual Demand of In-Vehicle Touchscreen Interactions}                      



%
\author[1]{Patrick Ebel}[orcid=0000-0001-7511-2910]
\cormark[1]
\ead{ebel@cs.uni-koeln.de}

\author[2]{Christoph Lingenfelder}[orcid=0000-0001-9417-5116]
\ead{christoph.lingenfelder@mercedes-benz.com}

\author[1]{Andreas Vogelsang}[orcid=0000-0003-1041-0815]
\ead{vogelsang@cs.uni-koeln.de}

\address[1]{University of Cologne, Cologne, Germany}

\address[2]{MBition GmbH, Berlin, Germany}

\cortext[cor1]{Corresponding author}



\begin{abstract}
With modern infotainment systems, drivers are increasingly tempted to engage in secondary tasks while driving. Since distracted driving is already one of the main causes of fatal accidents, in-vehicle touchscreen \acp{HMI} must be as little distracting as possible. To ensure that these systems are safe to use, they undergo elaborate and expensive empirical testing, requiring fully functional prototypes. Thus, early-stage methods informing designers about the implication their design may have on driver distraction are of great value. This paper presents a machine learning method that, based on anticipated usage scenarios, predicts the visual demand of in-vehicle touchscreen interactions and provides local and global explanations of the factors influencing drivers' visual attention allocation. The approach is based on large-scale natural driving data continuously collected from production line vehicles and employs the \ac{SHAP} method to provide explanations leveraging informed design decisions. Our approach is more accurate than related work and identifies interactions during which long glances occur with 68\,\% accuracy and predicts the total glance duration with a mean error of 2.4\,s. Our explanations replicate the results of various recent studies and provide fast and easily accessible insights into the effect of UI elements, driving automation, and vehicle speed on driver distraction. The system can not only help designers to evaluate current designs but also help them to better anticipate and understand the implications their design decisions might have on future designs.
\end{abstract}




\maketitle

 \section{Introduction}
 
\fix{Nowadays, large center stack touchscreens, like the ones found in Tesla's Model 3\footnote{\url{https://www.tesla.com/model3}} or the Mercedes-Benz EQS\footnote{\url{https://www.mercedes-benz.com/en/innovation/future-mobility/eqs-with-unique-mbux-hyperscreen/}} are the main interface between the driver and the \acp{IVIS}. During the interaction, drivers need to take their eyes off the road to scan the information presented on the screen.} Thus, they distribute their visual attention between the primary driving task and the secondary touchscreen task. This increases the risk of a crash significantly~\citep{Dingus.2016, Green.1999}, in particular for eyes-off-road durations longer than two seconds~\citep{Klauer.2006}.
With \acp{IVIS} becoming more complex and incorporating an ever-increasing amount of functionalities, drivers \fix{have more options than ever to interact with them while driving. The temptation to engage in non-driving related tasks is further increased by constantly improving driving automation features. During partially automated driving, drivers tend to engage more often in non-driving related tasks, even though they are still supposed to monitor the vehicle~\citep{Carsten.2012, Winter.2014, ebel.2022}.}
To ensure that \acp{IVIS} are safe to use, they are subject to strict regulations and elaborate test protocols. 
Automotive \acp{OEM} conduct expensive empirical user studies in artificial settings (e.g., driving simulators) to test the safety of the systems. \fix{However, driving simulator studies can only replicate real-world driving behavior to a certain degree~\citep{Riener.2011} and often lack absolute validity \citep{kaptein.1996, fisher.2011}. Furthermore, to evaluate a new \acp{IVIS} design in a user study, all relevant features need to be implemented in a functional prototype.}
Although such measures will remain necessary, automotive UX experts require explainable evaluation methods~\citep{ebel.RoleAndPotentials.2020} allowing them to identify potentially distracting interaction patterns already in the early design stages. Such automated methods can facilitate the development of interaction concepts that are safe by design and, therefore, less likely to fail final evaluations.

Current approaches that predict driver distraction based on user interaction information are derived from methods like Fitt's Law~\citep{Fitts.1954} or the \ac{KLM}~\citep{Card.1980}, where the time of certain operations is summed up to predict the total time of a task. The glance behavior is then derived from this metric. Although these models are limited by their cumulative linear nature and require experts to manually specify each task, they are highly interpretable. 
With more complex models, interpretability is often sacrificed for increased accuracy, highlighting the inherent trade-off between the two~\citep{Lundberg.2017}. \fix{However, without explanations, scientific findings may remain hidden, user acceptance suffers, and the learning effect is limited~\citep{Molnar.2020, Doshi.2017}.} To facilitate \fix{data-informed} design decisions, it is not only important to predict potentially dangerous interaction sequences but also to understand which interactions force drivers to take their eyes off the road.

\section{Background and Related Work}
In this section, we introduce the concept of visual demand. We explain how to measure it and present computational models of visual demand. Finally, we introduce \ac{SHAP}, an approach to generate explanations for machine learning predictions.

\subsection{Visual Demand of Secondary Task Engagements}

Visual demand is the \textit{``degree or quantity of visual activity required to extract information from an object to perform a specific task''} \citep{ISO15007}. While driving a car, visual distraction from the primary driving task by engaging in a secondary task, compromises driving performance and safety \citep{Engstroem.2005, Donmez.2010, Liang.2010, Green.1999, Klauer.2006, Horrey.2007}. This also applies to higher automation levels of driving automation \citep{SAEJ3016}. Recent research shows that takeover performance after a stretch of automated driving is significantly affected by the visual-cognitive load of the secondary task \citep{Wintersberger.2021} and distraction in general \citep{Merlhiot.2021}.
In ISO:15007:2020 \citep{ISO15007} multiple metrics to measure visual demand are described. Two of the metrics that are widely used are the \ac{TGD} and the average glance duration. The \ac{TGD} is the ``summation of all glance durations to an area of interest (or set of related areas of interest) during a condition task, subtask or sub-subtask''. The average glance duration is the ``mean duration of all glance durations to an area of interest (or set of related areas of interest) during a condition task, subtask or sub-subtask)''.
Further research \citep{Horrey.2007} shows that single longer-than-normal glances, especially those longer than two seconds \citep{Klauer.2006}, highly correlate with reduced driving safety. This is also mentioned in the ``Visual-Manual NHTSA Driver Distraction Guidelines for In-Vehicle Electronic Devices'' \citep{NHTSA.2012}. However, according to \citet{Victor.2014} there is not a single metric that can fully describe the relationship between glance behavior and risk, but rather a combination of metrics is necessary. \citet{Burns.2010} additionally argue that whereas any single measure only provides an incomplete assessment of distraction, empirically supported measures such as the above-introduced ones should be included for decision-making as early as possible in the design process.

\subsection{Visual Demand Prediction}

\fix{Various methods aim to predict visual-manual distraction while driving \citep{Kanaan.2019, Li.2018, Wollmer.2011, Li.2020, Risteska.2021}. Most of them focus on driver distraction detection to warn the driver when a potentially dangerous situation is detected.} These approaches are often based on naturalistic driving data and employ various machine learning methods. They utilize driving performance metrics (e.g., speed or steering wheel angle) \citep{Kanaan.2019, Li.2018, Wollmer.2011}, environmental data (e.g., traffic conditions) \citep{Risteska.2021}, or video data of the driver \citep{Li.2020, Kutila.2007}. While these approaches show promising results, they do not incorporate any information on how drivers interacted with secondary devices like mobile phones or \acp{IVIS}. Therefore, they do not generate insights into the visual demand of specific UI elements or interactions.

However, various approaches exist that model the visual demand of in-vehicle \acp{HMI} based on user interactions with specific UI elements. They explain the effect specific interactions have on drivers' visual distraction. These approaches focus on the understanding of interaction behavior and aim to identify distracting features of in-vehicle \acp{HMI}. Their purpose is to inform designers and researchers in the early stages of the development process about possible implications their design might have on driver distraction. In this work, we focus on the latter and provide an overview of the current state-of-the-art in this domain.

Most of the approaches that predict visual demand, based on user interactions, are bottom-up approaches derived from the \ac{KLM} modeling technique \citep{Card.1980, Card.1983}. In such approaches, an entire task is decomposed into a sequence of specific primitive operators (e.g., pressing a button, orsearching in a list). The interaction durations for each operator are then determined empirically \citep{Schneegass.2011}. The overall time on task predictions are then equal to the sum of the individual interaction durations of the respective operators occurring in the task. The \ac{KLM} technique was originally developed to predict processing times in computer-assisted office work, but multiple adjustments were made to assess \acp{IVIS} \citep{Schneegass.2011, Manes.1997, Lee.2019}. However, most of these approaches focus on task completion times rather than visual demand. \citet{Pettitt.2007} were the first to propose a \ac{KLM}-based approach to predict visual demand. They show a high correlation between predicted values and measures from an occlusion experiment. The first KLM-based method to directly predict visual demand is presented by \citet{Purucker.2017}, who propose a task-specific \ac{KLM} model. They argue that using fixed operators to model innovative and new hardware can only work to a limited extent. Whereas their approach can only predict \ac{TGD}, \citet{Large.2017} propose a method that can additionally predict the number of glances and the mean glance duration. Their information-theoretic approach is based on the Hick-Hyman Law for decision/search time and Fitt's Law for pointing time.

Whereas the results achieved by the presented KLM-based approaches are promising, they all share several drawbacks.
First, due to their cumulative and linear character, the models are not suited to model potential (non-linear) dependencies between different user interactions or driving situations. For example, the difference in the visual demand between selecting an element out of a list and tapping a button might be negligible for lower speeds but significant for higher speeds. Additionally, the length of an interaction sequence in combination with specific interactions might also influence visual demand in a non-linear and non-additive way \citep{Purucker.2017}. For example, if the driver presses two buttons that are located close to each other, it unlikely results in a doubling of the \ac{TGD} as the driver might perform both interactions during one glance. 
Second, the model parameters of the introduced approaches are derived from empirical testing in restricted driving scenarios using driving simulators of different fidelity, and a relatively small number of participants. This can likely lead to predictions being very context-dependent, as also noted by \citet{Large.2017} and shown in a real-world driving experiment evaluating the applicability of Fitt's Law \citep{Pampel.2019}. 
Third, current approaches do not consider the effect the driving situation has on the visual demand. Research shows that drivers modulate their task engagement and visual attention based on driving demands \citep{Risteska.2021} and the degree of assisted driving \citep{Morando.2021, Tsimhoni.2001, Gaspar.2019, Large.2017a} making it important to include such parameters.

A different approach is taken by \citet{Kujala.2015} who propose a model based on the ACT-R cognitive model architecture \citep{Anderson.2004}. Their approach aims to represent the visual sampling strategy of drivers. They argue that drivers adjust their glances based on a time limit that's dependent on the current driving performance. Whereas the model can predict multiple facets of visual demand, only grid and list layouts are considered. Furthermore, the driving scenario is fairly simple and the evaluation shows significant drawbacks in prediction accuracy, especially concerning the detection of long glances.

We argue that the utilization of large natural driving and interaction data in combination with machine learning approaches that can model non-linear relationships can be a promising step toward more accurate and holistically applicable solutions.
Machine learning approaches have led to high-quality predictions in the domain of distraction detection methods as introduced above.

However, two main factors prevent these approaches from generating valuable insights into how specific design elements affect visual demand. First, interaction data is not yet available in a similar quantity as driving data. Second, all the above-presented machine learning approaches lack explainability. Whereas certain performance metrics are reported, the models remain a black box without providing insights on the features that are decisive for the predictions.

\subsection{Explainable Predictions with SHAP}

\ac{XAI} aims to make machine learning models more transparent by providing human-understandable (interpretable) information, explaining the behavior and processes of machine learning models \citep{barredoarrieta.2020, liao.2020}. Explanations serve as an interface between the human and the model \citep{barredoarrieta.2020} and can be valuable in various applications \citep{wiegand.2020, wang.2021}.
Explanations can enhance scientific understanding \citep{Doshi.2017}, increase user trust \citep{shin.2021, Lipton.2018}, and can help to infer causal relations in data \citep{Verma.2020}. For the task at hand explainable predictions are of particular interest because the goal is not only to make predictions of the visual demand but also to draw conclusions about the impact of specific UI elements, gestures, and varying driving situations. The goal of this approach is to enable AI-assisted decision-making \citep{zhang.2020}, optimizing a joint decision based on the domain knowledge of the human expert and the insights generated by the model prediction and accompanying explanation.

\ac{SHAP}, proposed by \citet{Lundberg.2017} is a method based on Shapley values from coalitional game theory \citep{Shapley.1953}. The \ac{SHAP} method provides local and global explanations for arbitrary predictive models. \ac{SHAP} belongs to the class of additive feature attribution methods. The main idea is to use an interpretable explanation model $g(z')$ in the form of a linear function such that the model's prediction of a certain instance is equal to the sum of its feature contributions $\phi_{i} \in \mathbb{R}$ \citep{Molnar.2020}:

\begin{equation}\label{eq:AdditiveMethods}
\centering
g(z') = \phi_{0} + \sum_{i=1}^{M} \phi_{i} z'_{i}\,,
\end{equation}

where $z' \in \{0,1\}^M$ with $z'_{i}$ represents the presence of feature $i$, $\phi_0$ represents the models output in case no feature is present, and $M$ is the number of input features \citep{Lundberg.2020}.

\citet{Lundberg.2017} further state that a single unique solution exists that follows the definition of additive feature attribution methods (see \ref{eq:AdditiveMethods}) and satisfies the properties of local accuracy, missingness, and consistency. Local accuracy describes that the sum of the feature attributions is equal to the prediction of the original model. Missingness describes that a missing feature ($z_i = 0$) gets assigned an attribution of zero and consistency states that when changing a model such that it is more dependent on a certain feature, the attribution of that feature should not decrease.

The only possible solution as described by \citet{Lundberg.2017} is given by the \ac{SHAP} values:
\begin{equation}\label{eq:Haversine1}
\centering
\phi_i=\sum_{S\subseteq N\setminus\{i\}} \frac{|S|!(|M|-|S|-1)!}{M!}  \left(f_x(S\cup\{i\})-f_x(S)\right)\,,
\end{equation}

with $S$ being the set of non-zero indexes in $z'$, $f_x(S)$ being the expected value of the function conditioned on a subset $S$ of the input features, and $N$ being the set of all input features.

Multiple different approaches exist to approximate SHAP values for different kinds of machine learning models. However, in this study, we use TreeSHAP \citep{Lundberg.2020} which allows the computation of exact SHAP values for tree-based approaches.

Compared to approaches like LIME \citep{Ribeiro.2016} or approaches specific to tree-based models like permutation importance of feature impurity calculations, \ac{SHAP} has many advantages. Due to the solid foundation in game theory \citep{Molnar.2020}, \ac{SHAP} values come with theoretical guarantees about consistency and local accuracy. Additionally, local and global explanations are consistent, \ac{SHAP} values indicate whether the contribution of each feature is positive or negative, and \citet{Lundberg.2018} show a greater overlap of \ac{SHAP} values and human intuition \citep{Lundberg.2017, Lundberg.2020}.

\section{Proposed Approach}

In this work, we showcase how to effectively use machine learning methods to predict and explain the visual demand of in-vehicle touchscreen interactions based on large naturalistic driving data. The contribution of this paper is two-fold: First, we propose a machine learning approach predicting the visual demand of in-vehicle touchscreen interactions based on the type of interaction and the associated driving parameters. Second, we apply the \ac{SHAP} method~\citep{Lundberg.2017} to explain the predictions and to visualize how user interactions, vehicle speed, steering wheel angle, and automation level, affect drivers' long glance probability and \ac{TGD}.
In the following, we introduce several definitions that will be used in the remainder of the paper. Furthermore, we describe the data collection, data processing, and modeling procedures in detail.

\subsection{Definitions and Problem Statement}\label{ch:definition}
The goal of this approach is to predict drivers' visual attention allocation based on user interactions and the associated driving parameters. To do so, we model drivers' secondary task engagements by combining interaction sequences, driving sequences, and glance sequences. These concepts are introduced in the following.\footnote{For the formal definitions refer \ref{ch:AppendixDefinitions}}

\textbf{Interaction Sequence.} An interaction sequence is defined as a set of subsequent touchscreen interactions performed by the driver. The duration between two subsequent interactions must be smaller than $\Delta t_{max}$.

\textbf{Glance Sequence.} A glance sequence is defined as a set of subsequent driver glances toward a predefined \ac{AOI}.

\textbf{Driving Sequence.} A driving sequence is a sequence of driving data observations. Each observation consists of the vehicle speed, the steering wheel angle, and the status of \ac{ACC} and \ac{SA}.

\textbf{Secondary Task Engagement.} A secondary task engagement $S$ describes the touchscreen interactions, the driving behavior, and the glance behavior of a driver while interacting with the center stack touchscreen. We consider the vehicle speed and steering wheel angle from $t_b$ seconds before the first until $t_b$ seconds after the last interaction. Furthermore, all glances that fall in between the first and last interaction are considered.

\textbf{Long Glance Prediction Task.} The long glance prediction task describes the task of predicting if the driver will look at the center stack touchscreen for more than two seconds during a secondary task engagement.

\textbf{\ac{TGD} Prediction Task.} The \ac{TGD} prediction task describes the task of predicting the drivers' total glance duration toward the center stack touchscreen during a secondary task engagement.

\subsection{Data Collection}

The data used in this study was collected over the air from over 100 test cars and five different car models \fix{via the Mercedes-Benz telematics data logging framework. A more detailed description of the logging mechanism is provided by \citet{Ebel.2021}}. The data collection period ranged from mid-October 2021 to mid-January 2022. All company internal test cars with the latest software architecture, an eye-tracker, and ACC and SA functionality, contributed to the data collection. The test cars were used for various, mostly not UI-related test drives, but also for leisure drives of employees. While the cars were driven all over Europe, most of the trips were recorded in Germany. We leveraged the data collection and processing framework of Mercedes-Benz, to collect touchscreen interactions, driving data (vehicle speed, steering wheel angle, and level of driving automation), and eye-tracking data. We did not collect any demographic or environmental data.

The touchscreen interactions were collected from the UI software, where a datapoint was logged each time the driver touched the center stack touchscreen. A datapoint consists of the touched UI element, the start and end position of all fingers used for the gesture, and a timestamp.

The glance data was acquired via a stereo camera located in the instrument cluster behind the steering wheel. The system is already commercialized and available in production line vehicles. As with most remote eye trackers, gaze detection is based on the pupil-corneal reflection technique~\citep{merchant.1967}. In this method, the pupil center is tracked in relation to the position of the corneal reflection~\citep{hutchinson.1989}. The driver’s field of view is divided into different \acp{AOI} and a new glance is collected each time the focus of the driver switches between \acp{AOI}. Across all conditions, the average true positive rate for each of the \acp{AOI} is above 90 percent. The system captures no raw video data.

All driving-related data was directly collected from the \ac{CAN} bus. The continuous signals (vehicle speed and steering wheel angle) were collected with a frequency $4\,Hz$, and the event-based signals (\ac{ACC}, \ac{SA}, and seat belt information) were collected on change. \ac{ACC} automatically adjusts the vehicle speed based on speed limits and the vehicles ahead. \ac{SA} actively supports lateral control to keep the car centered in the lane. Both systems operate at speeds between 0\,km/h and 210\,km/h.

\subsection{Data Processing}
\begin{figure*}[pos=htpb!, align=\centering, width=\linewidth]
		\includegraphics[width=\linewidth]{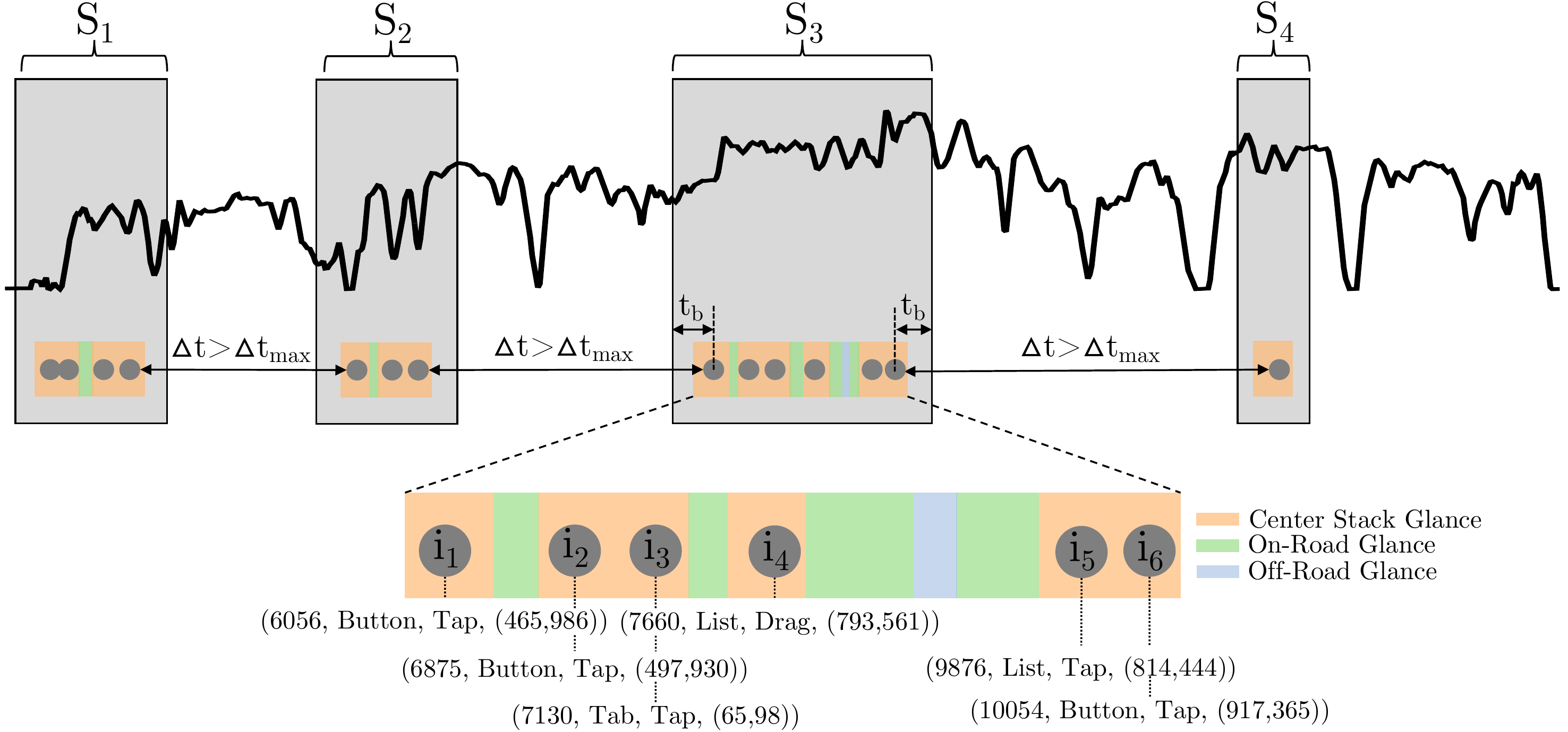} 
		\caption{\fix{Schematic overview on how secondary task engagements $S_n$ are extracted from driving sequences (solid black line), glance sequences (colored rectangles), and interaction sequences (grey dots).}}
		\label{fig:SequenceSketch} 
	\end{figure*}

In the following we describe how we process the data such that the respective sequences follow the definitions given in \ref{ch:definition}. Fig. \ref{fig:SequenceSketch} shows a schematic overview describing the data synthesis of the individual sequences. For clarity, we only display the vehicle speed \fix{(solid black line)}, representative of the other driving data.

\subsubsection{Interaction Data}
In controlled experiments, participants are instructed to perform pre-defined tasks that are specified by the experimenter. In such settings, it is straightforward to map user interactions and tasks. However, in the observational setting at hand, no task boundaries are defined. We do not know which task the driver intended to complete during the secondary task engagement, nor are the start or end times defined. One possibility to extract interaction sequences would be to consider all interactions that occurred during one trip. However, this would lead to very long interaction sequences with dense clusters of interactions sparsely distributed over a long period of time. To solve this problem, we set the maximum update interval, as defined in \ref{ch:definition} to $10\,s$, $\Delta_{t_{max}} = 10\,s$. We assume that after a period of 10 seconds with no interaction the driver disengaged from the secondary task and the interaction sequence ended after the last interaction. We then consider the next interaction as the starting point of a new interaction sequence. We argue that the 10-second assumption is valid, because both the distribution of interaction sequence durations and the distribution of total glance times toward the center stack touchscreen match well with values reported in the literature \citep{NHTSA.2012, Angell.2008}.

\begin{table}[pos=htpb!, align=\centering]
	\small
	\centering
	\caption{Overview of the final input features describing a secondary task engagement.}
	\label{tab:InputFeatures}
		\begin{tabular}{ll}
		    \toprule
			Feature & Description \\
			\toprule
			\multicolumn{2}{l}{Interaction Data}\\
			$n_{\text{Button}}$         & $\#$ Interactions with regular buttons (e.g., push or radio buttons)\\
			$n_{\text{List}}$           & $\#$ Interactions with lists (e.g., when choosing a suggested destination)\\
			$n_{\text{Map}}$           & $\#$ Interactions with a map viewer (e.g., when zooming or dragging the navigation map)\\
			$n_{\text{Slider}}$         & $\#$ Interactions with slider elements (e.g., when changing the volume) \\
			$n_{\text{Homebar}}$        & $\#$ Interactions with the static homebar on the bottom of the screen \\
			$n_{\text{CoverFlow}}$      & $\#$ Interactions with cover flow widgets (e.g., when scrolling through albums covers) \\
			$n_{\text{AppIcon}}$        & $\#$ Interactions with app icons on the home screen \\
			$n_{\text{Tab}}$            & $\#$ Interactions with tab bars \\
			$n_{\text{Keyboard}}$       & $\#$ Interactions on the keyboard or number pad (e.g., when entering a destination) \\
			$n_{\text{Browser}}$        & $\#$ Interactions within the web browser \\
			$n_{\text{RemoteUI}}$       & $\#$ Interactions within Apple Car Play or Android Auto \\
			$n_{\text{ControlBar}}$     & $\#$ Interactions with a control bar, displayed as a small overlay on various screens\\
			$n_{\text{PopUp}}$          & $\#$ Interactions with pop-up element\\
			$n_{\text{ClickGuard}}$     & $\#$ Interactions with non-interactable background elements \\
			$n_{\text{Other}}$          & $\#$ Interactions with a UI element that does not fit any of the above categories \\
			$n_{\text{Unknown}}$        & $\#$ Interactions with a UI element for which the identifier could not be extracted \\
			$n_{\text{Tap}}$            & $\#$ Tap gestures\\
			$n_{\text{Drag}}$           & $\#$ Drag gestures\\
			$n_{\text{Multitouch}}$     & $\#$ Multitouch gestures \\
			$d_{\text{avg}}$            & Average distance between two consecutive interactions in \textit{px} \\
			$N$                         & Number of interactions\\
			\midrule
			\multicolumn{2}{l}{Driving Data}\\
			$v_{\text{avg}}$                   & Average vehicle speed in \textit{km/h}\\
			$\theta_{\text{avg}}$              & Average steering wheel angle in \textdegree\\
			$a_{\text{acc}}$            & Status of the adaptive cruise control $a_{\text{acc}} \in \{0,1\}$\\
			$a_{\text{sa}}$             & Status of the steering assist $a_{\text{sa}} \in \{0,1\}$
		\end{tabular}
\end{table}

For all remaining interactions, we compute the gesture type (\textit{Tap, Drag, and Multitouch}) and distance between two interactions from the positioning information of the fingers. Finally, we map each UI element to an overarching element type (see Table \ref{tab:InputFeatures}). This step reduces data sparsity and ensures that the approach can produce generalizable statements about specific element types.

\subsubsection{Glance Data}
We apply multiple filtering steps to improve the data quality of the glance data. The filtering is partially adapted from related research \citep{Morando.2019}. In the first step, the glance information is aggregated into broader \acp{AOI} (\textit{On-road, Off-road, Center Stack}). \fix{According to ISO 15007-1:2020 \citep{ISO15007}, we consider all glances that are not directly directed on the road (e.g., glances in the rear-view mirrors) as off-road glances.} As we are explicitly interested in glances toward the center stack touchscreen, we distinguish these glances from regular off-road glances. Second, as described in Section \ref{ch:definition}, we consider all glances between the first and last touchscreen interaction of the associated interaction sequence. \fix{Fragmented glances at the beginning or end of an interaction sequence that start before the first interaction or end after the last interaction are considered as a whole.} Third, short periods (less than $300\,ms$) of tracking loss are interpolated if the \ac{AOI} preceding the loss is equal to the one succeeding it. Loss of tracking can occur due to changing lighting conditions, reflections in glasses, or when the camera view is blocked by the driver's hands on the steering wheel. Fourth, according to ISO 15007-1:2020 \citep{ISO15007} glances shorter than $120\,ms$ are interpolated because shorter fixations to an area of interest are physically not possible. Fifth, following the same argumentation, loss of tracking between different \acp{AOI} shorter than $120\,ms$ is interpolated as well. Sixth, eye-lid closures shorter than $500\, ms$ are interpolated to remove blinks as suggested by ISO 15007-1:2020 \citep{ISO15007}.

\subsubsection{Driving Data}
The driving data consists of continuous data and event-based data. In the first step, we extract the data that is relevant for the associated interaction sequences. To get a more stable estimate of the driving parameters in case of very short interaction sequences that might only consist of a single interaction, we consider steering wheel and vehicle speed data starting two seconds before the first interaction until two seconds after the last interaction of an interaction sequence ($t_b = 2\,s$). After data extraction, sequences with missing values or error values, and sequences that show deviations in the logging frequency are discarded.

\subsubsection{Data Aggregation and Final Filtering}
In total, we extracted 322,425 touchscreen sequences. We obtained valid speed data for 145,973 sequences, valid steering data for 81,150 sequences, and valid glance data for 111,792 sequences. After individual processing, we computed the intersection of the individual data sources resulting in 30,158 complete secondary task engagements. \fix{Most of the sequences were excluded either because they were generated on a test bench (no driving data was available), the car wasn't equipped with a camera, or because the sampling requirements were violated due to a loss of data connection.} In the second stage of data processing, we apply further filtering steps to increase data quality. \fix{To prevent the data from becoming too sparse, we} discarded 342 secondary task engagements with more than 41 interactions ($N>41$  corresponds to the $99^{th}$ percentile of the distribution of $N$). \fix{These secondary task engagements can be considered outliers without providing additional benefits for the use case at hand.} We further discard 16,864 secondary task engagements where a passenger was present \fix{because they also tend to interact with the center stack touchscreen. These interactions can not be mapped to driver glances and would skew the data toward fewer and shorter glances per secondary task engagement with many interactions logged that did not originate from the driver. Furthermore, we discarded }809 engagements during which the car came to a full stop and one sequence due to a remaining speed error. After this processing step, we obtained the final set of 12,142 secondary task engagements. Finally, we compute summary statistics for the secondary task engagements to generate the final set of features as described in Table \ref{tab:InputFeatures}. These features serve as input to the models introduced in the following.

\subsection{Modeling}
As formulated in the problem statement we solve one classification task (long glance prediction) and one regression task (TGD prediction).
For each of the tasks, we compare a \textit{Baseline} approach and a \textit{Logisitc/Linear Regression} approach against three machine learning approaches, namely \textit{Random Forests}, \textit{Gradient Boosting Trees}, and \textit{\acp{FNN}}. In the long glance prediction task, the \textit{Baseline} approach randomly predicts one of the two classes (i.e. in a balanced dataset the probability of correctly predicting a long glance is roughly 50\%). In the \ac{TGD} prediction task, the baseline approach predicts the median \ac{TGD} of the training dataset. The parameters of the machine learning-based methods are chosen based on extensive hyperparameter optimization using random search\footnote{For more details on the search space refer to: \ref{ch:AppendixHyperOpt}}.

\section{Evaluation}
In this section we present the final dataset, put the experimental results in perspective, and elaborate on the explainable predictions generated by applying the \ac{SHAP} method.

\subsection{Dataset}
\begin{figure}[pos=htpb!, , align=\centering, width=\linewidth]
	\captionsetup{width=\linewidth}
	\subfloat[Vehicle speed\label{subfig:avg_speed}]{%
		\includegraphics[width=0.325\linewidth]{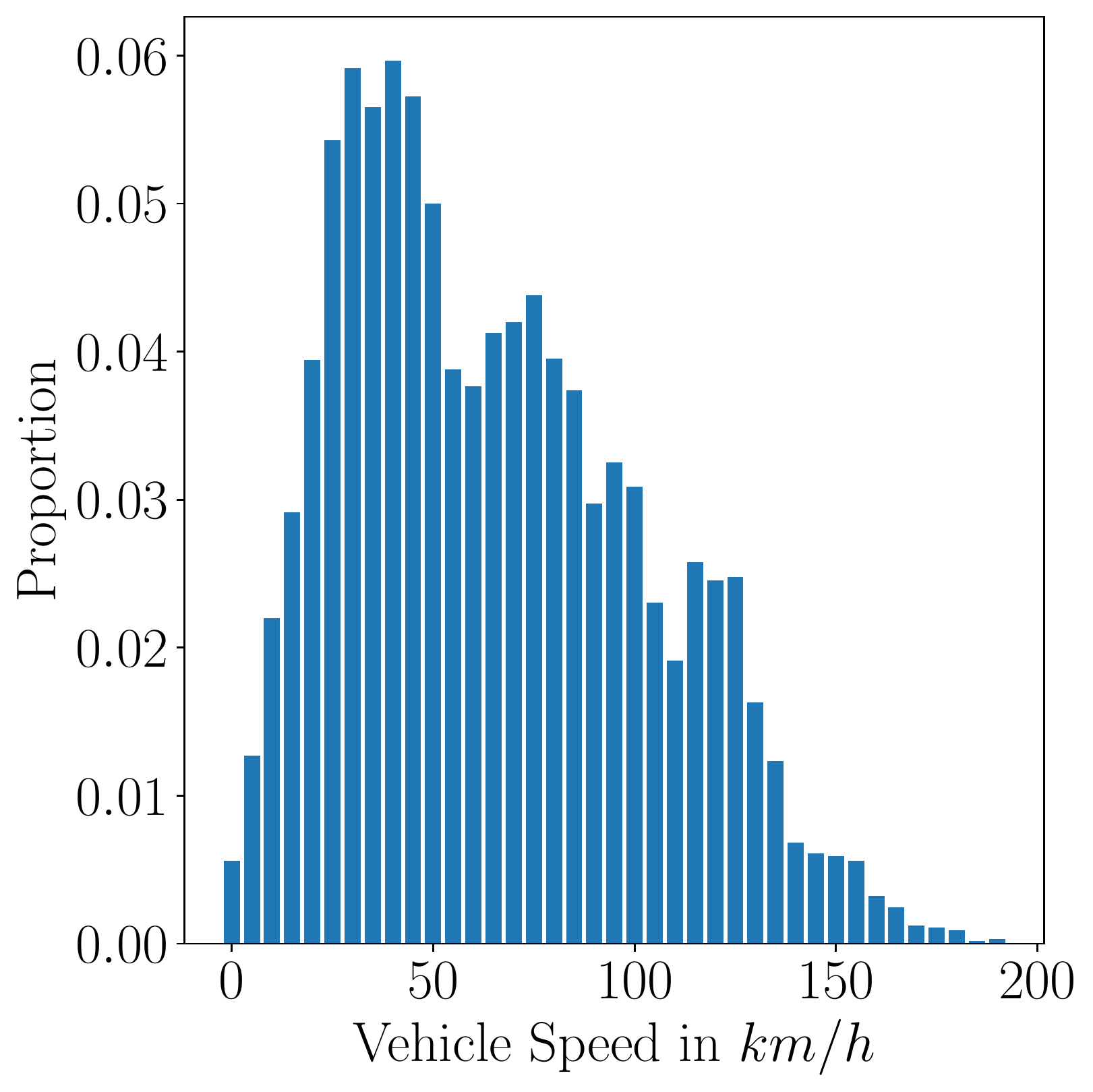}
	}
	\hfill
	\subfloat[Number of interactions\label{subfig:num_interactions}]{%
		\includegraphics[width=0.325\linewidth]{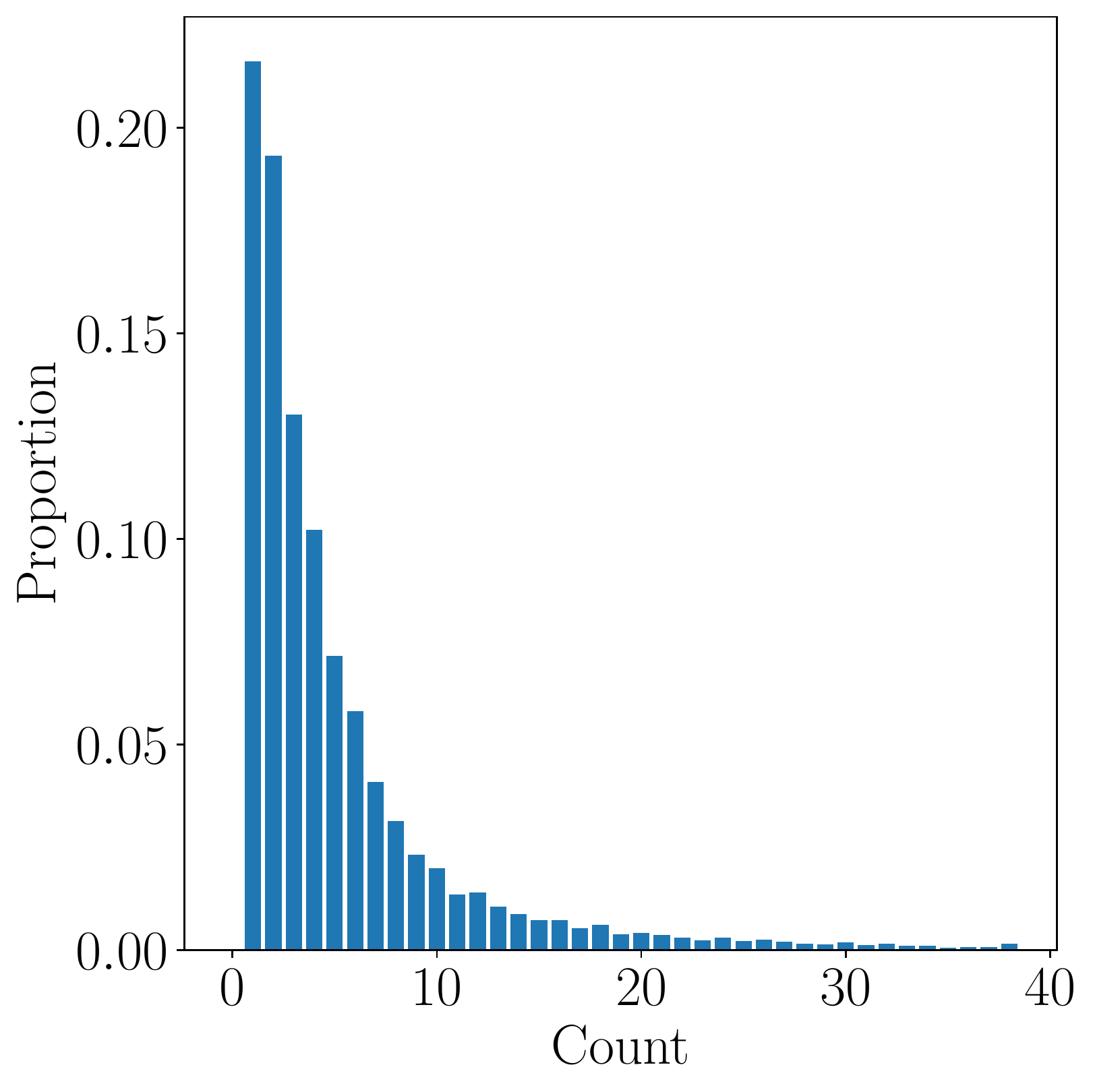}
	}
	\hfill
	\subfloat[Sequence duration\label{subfig:duration_sequence}]{%
		\includegraphics[width=0.325\linewidth]{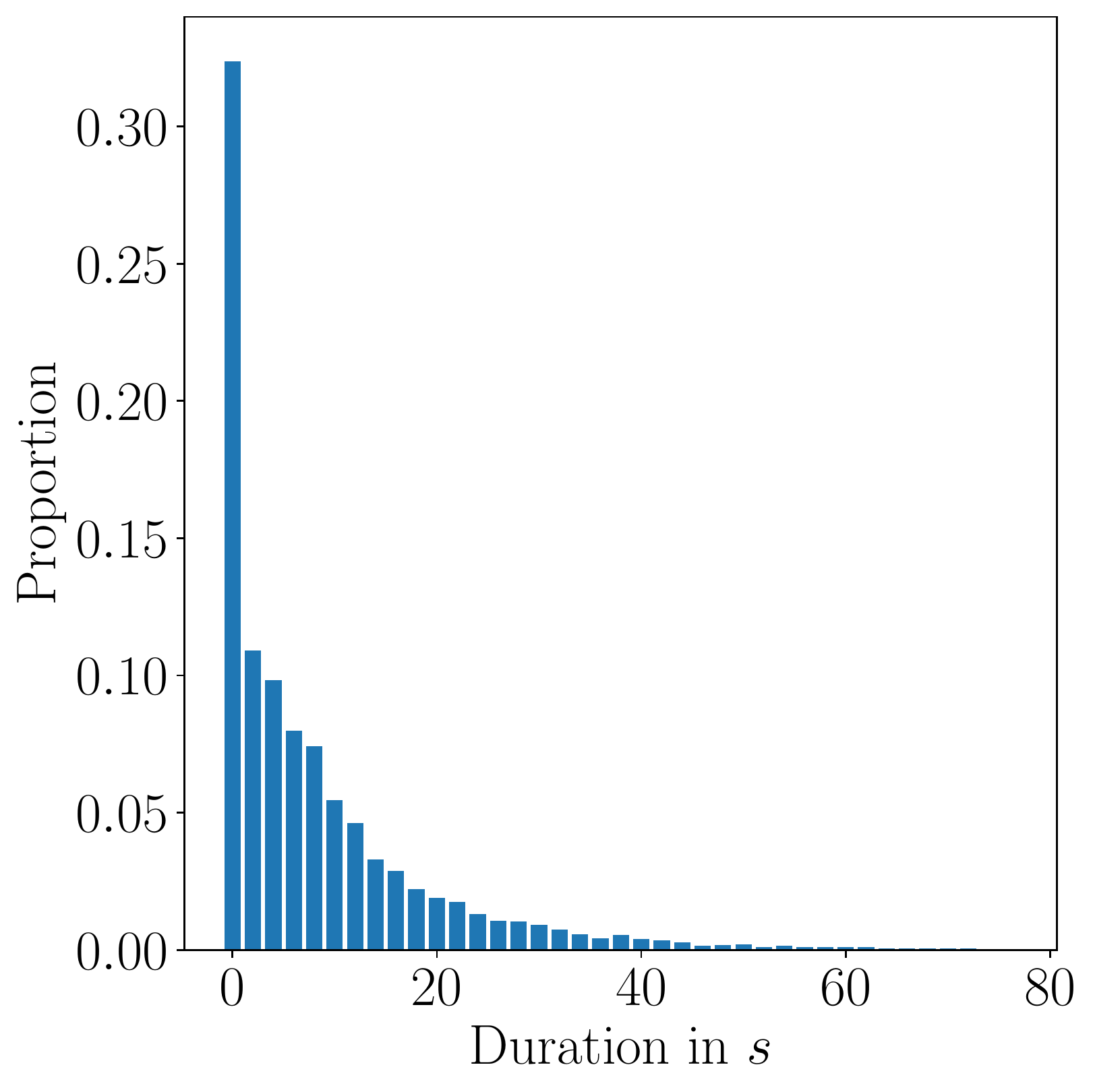}
	}	
	\hfill
	\subfloat[Number of glances\label{subfig:num_glances_hu}]{%
		\includegraphics[width=0.325\linewidth]{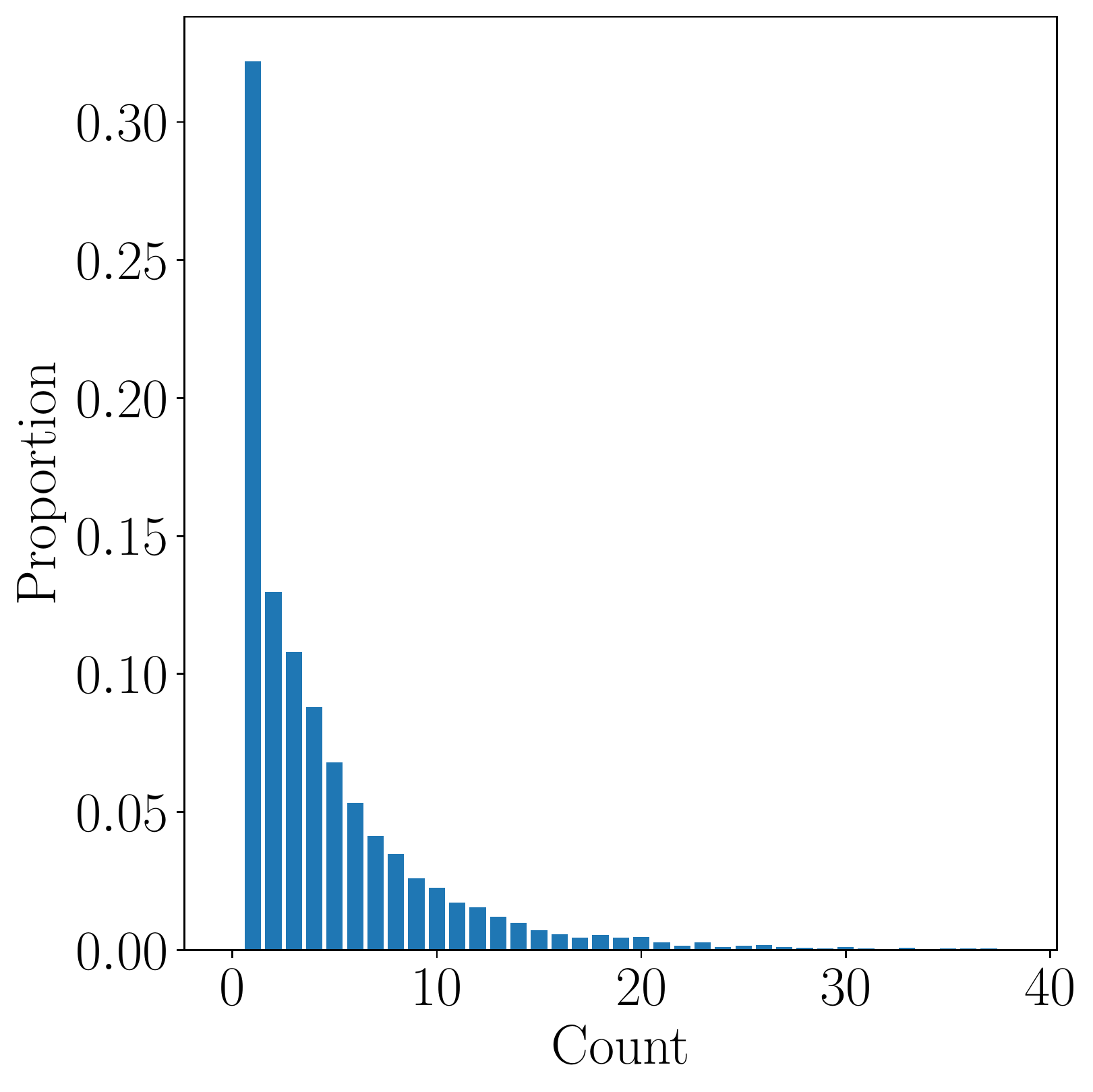}
	}
	\hfill
	\subfloat[On-road glances\label{subfig:glance_duration_on_road}]{%
		\includegraphics[width=0.325\linewidth]{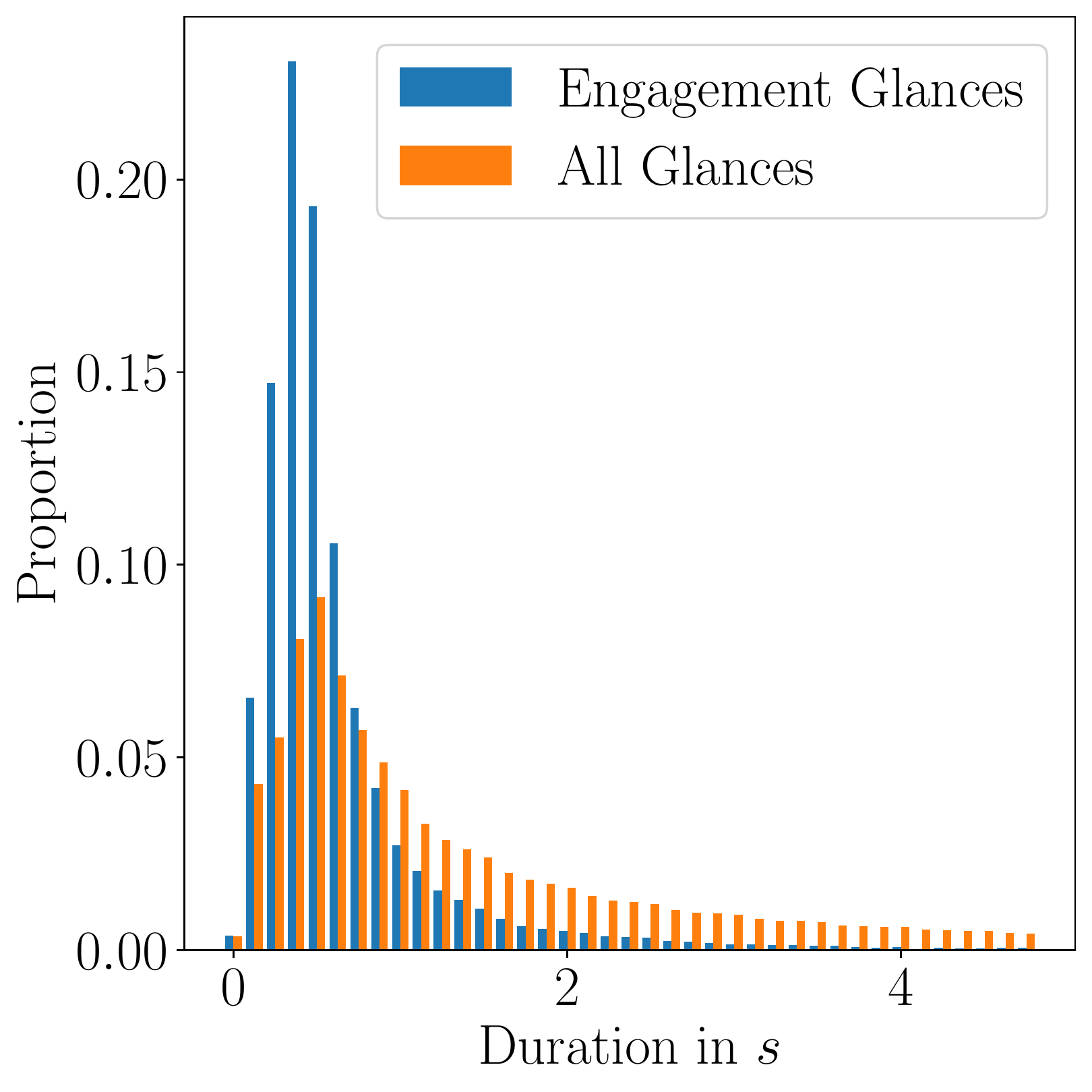}
	}
	\hfill
	\subfloat[Center stack glances \label{subfig:glance_duration_hu}]{%
		\includegraphics[width=0.325\linewidth]{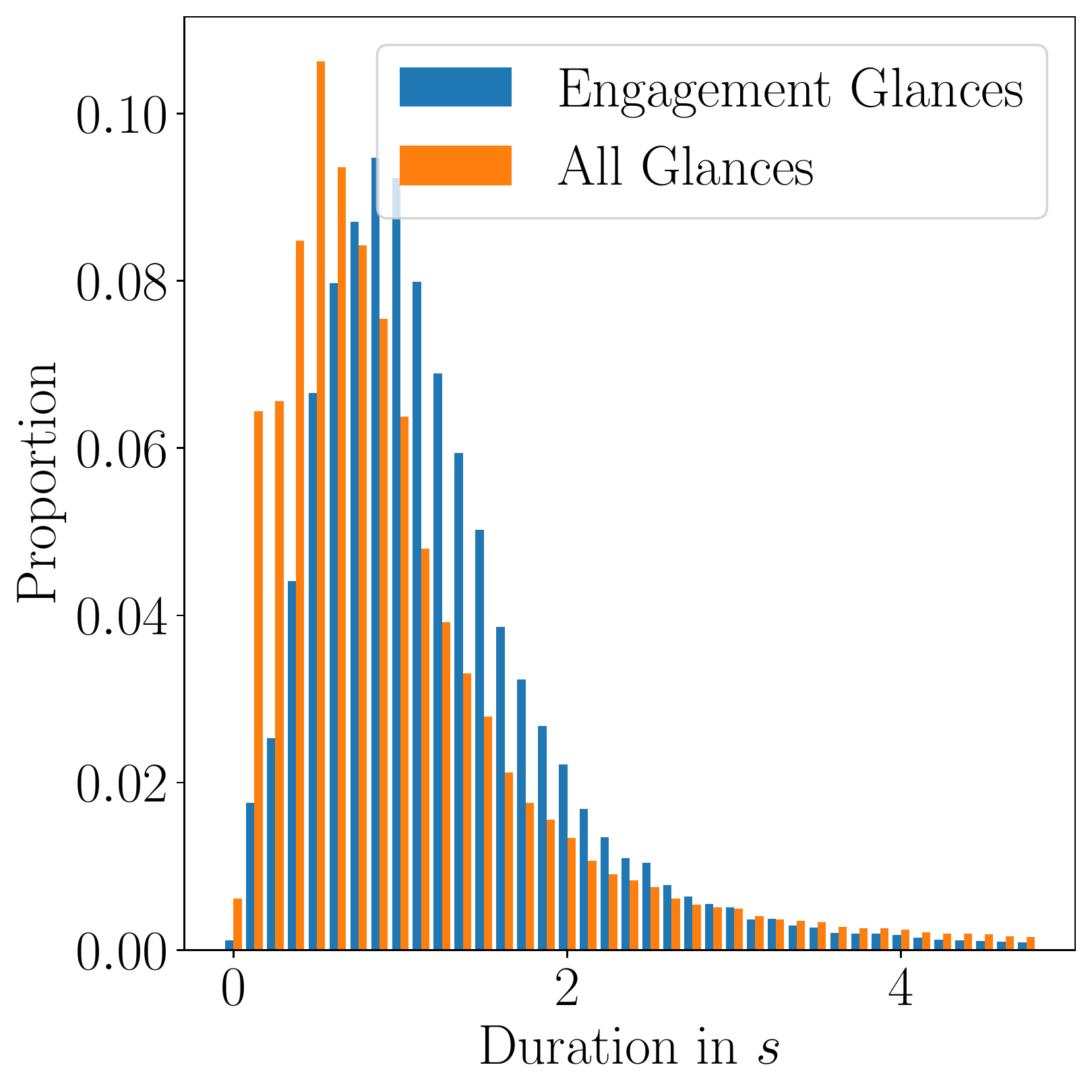}
	}
		\hfill
	\subfloat[TGD center stack\label{subfig:total_glance_duration_hu}]{%
		\includegraphics[width=0.325\linewidth]{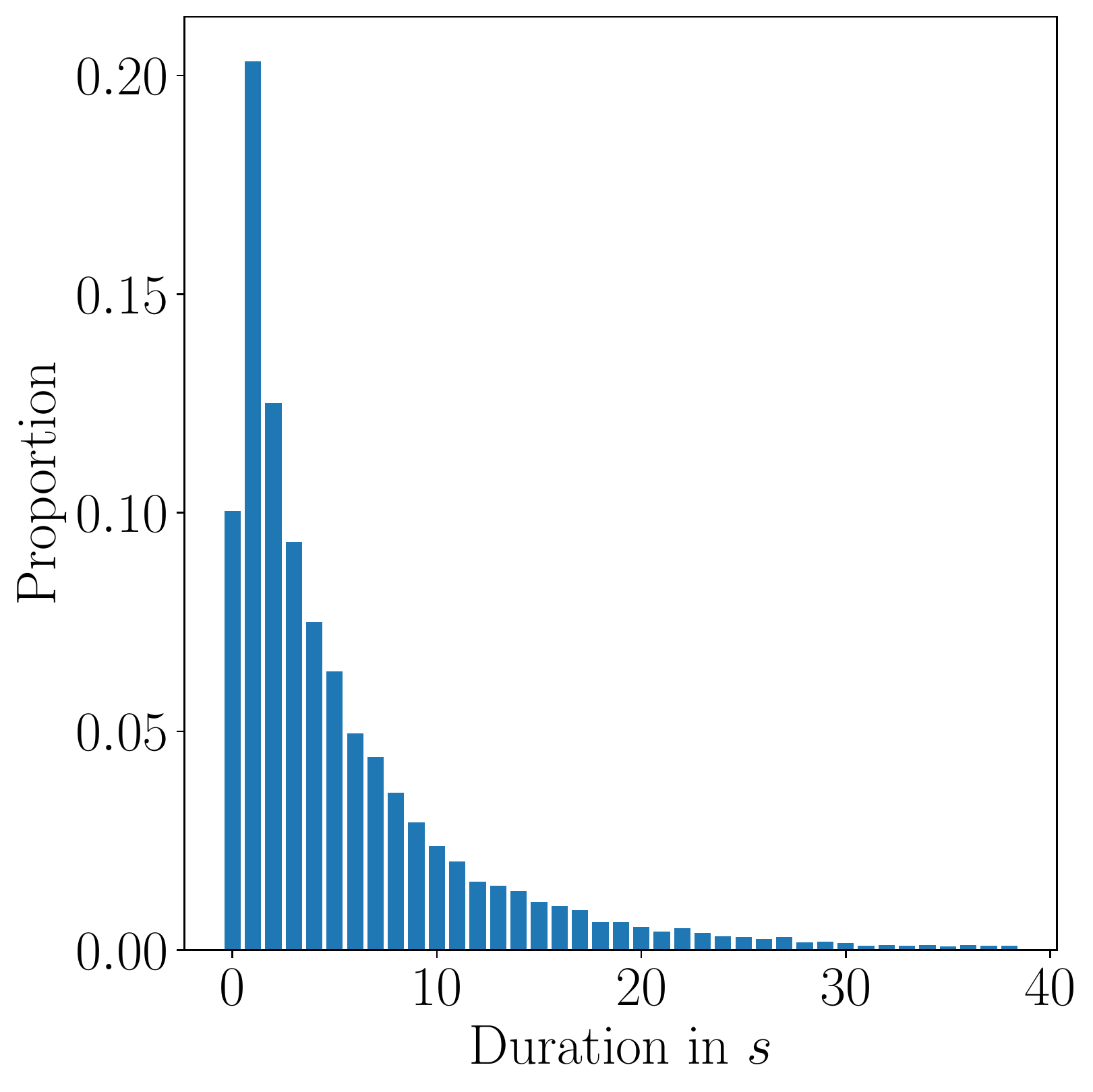}
	}
		\hfill
	\subfloat[Off-road glances\label{subfig:glance_duration_off_road_kde}]{%
		\includegraphics[width=0.325\linewidth]{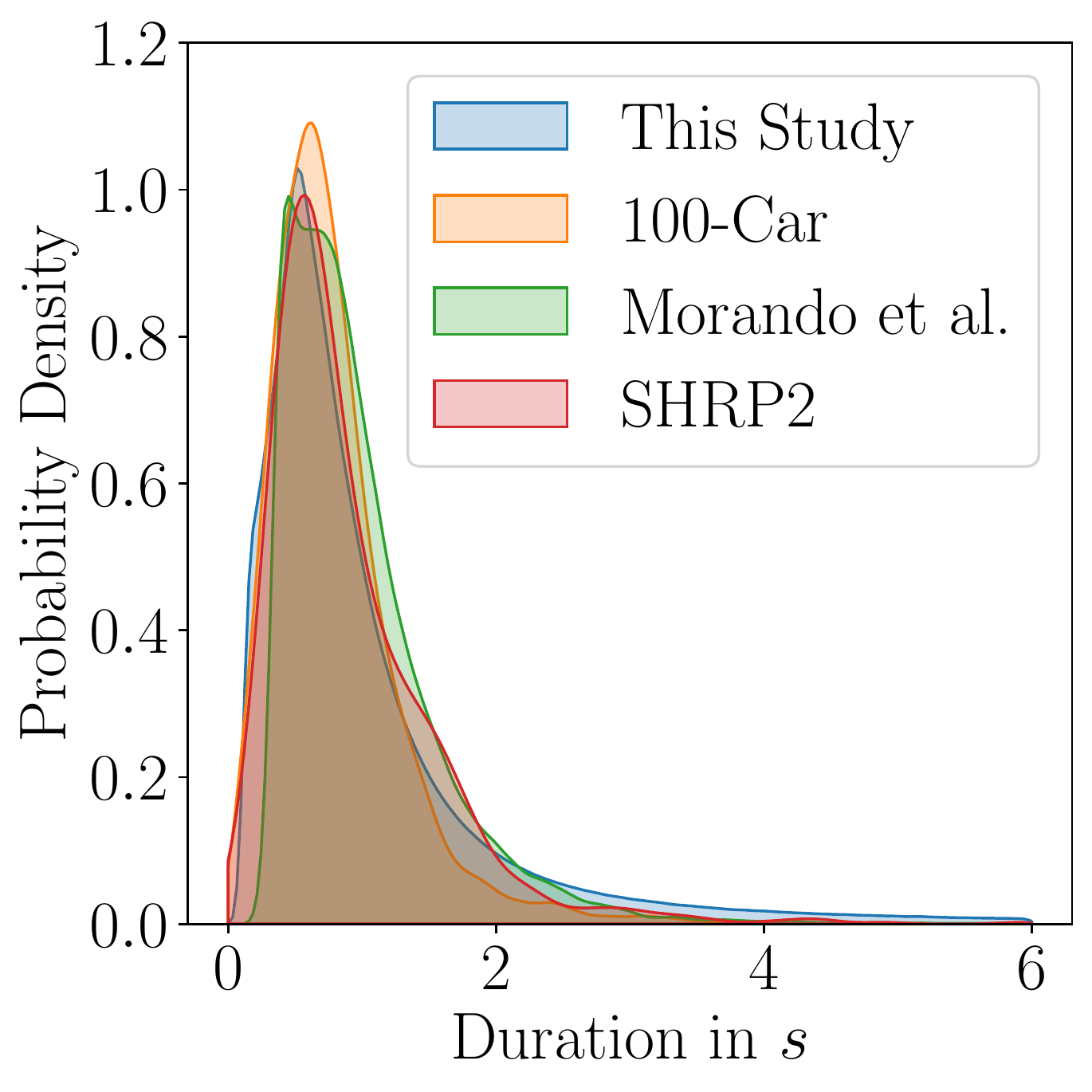}

	}
		\hfill
	\subfloat[On-road glances\label{subfig:glance_duration_on_road_kde}]{%
		\includegraphics[width=0.325\linewidth]{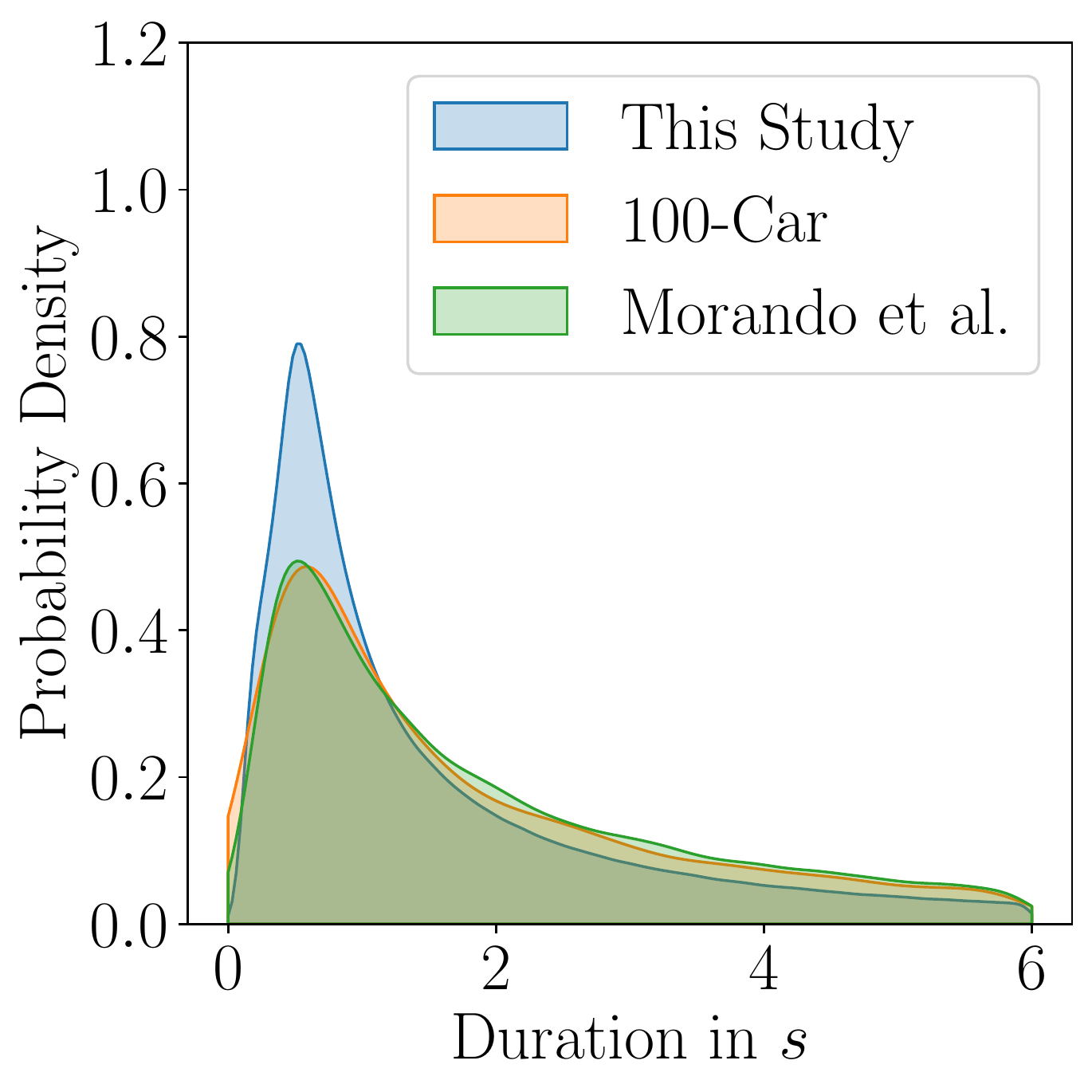}
	}
	
	\caption{Histograms that visualize the glance and interaction data. They show the (a) average speed per secondary task engagement,  (b) the number of interactions per secondary task engagement, (c) the duration of the interaction sequences of a secondary task engagement, and the (d) number of glances toward the touchscreen during per secondary task engagement. Figure (e) compares the on-road glance duration distribution for glances during a secondary task engagement with glances irrespective whether a touchscreen interaction was performed or not. Figure (f) establishes the same comparison for glances toward the center stack touchscreen. \fix{Figure (g) shows the total glance duration toward the center stack touchscreen during a secondary task engagement. Figure (h) and (i) compare the probability density functions of the off-road and on-road glance duration from this study with the 100-Car study, the SHRP2 study (only off-road glances), and the study of \citet{Morando.2019}.}
	\label{fig:Dataset}}
\end{figure}

The final dataset consists of 12,142 secondary task engagements sampled from 3,046 individual trips. The majority of secondary task engagements were collected from the Mercedes-Benz S-Class (7,342 secondary task engagements), EQS (3604), and EQE (824) models. \fix{The cars were equipped with a 17.7", 12.8", or 11.9" center stack touchscreen with similar pixel density.} In total, 61,943 touch interactions and 119,770 individual glances were collected. The median trip length is 34.28 minutes ($Q_1 = 17.49, Q_3 = 66.58$).
Specific glance and interaction statistics of the final dataset are presented in Figure~\ref{fig:Dataset}.\footnote{For the full dataset statistics refer to \ref{ch:AppendixSummaryStats}}

In Figure~\ref{subfig:glance_duration_on_road} and Figure~\ref{subfig:glance_duration_hu}, the glance duration distribution during secondary task engagements (blue) is plotted against the glance duration distribution over all sessions independent of the driver being engaged in a secondary task (orange). This allows a comparison with approaches that utilize data collected irrespective if the driver being engaged in a secondary task or not.

\fix{We further compare our data with the manual driving baseline of the \textit{100-Car Study}~\citep{Dingus.2006} (data provided by \citet{custer.2018}, the \textit{SHRP2}~\citep{Victor.2014} (data available in \citet{bargman.2015}) and the data reported in the work of \citet{Morando.2019} (provided by the authors upon request). Figure~\ref{subfig:glance_duration_on_road_kde} and Figure~\ref{subfig:glance_duration_off_road_kde} show the glance distribution of on-road and off-road glances for the respective datasets. The glance distributions were truncated at 6 seconds since this corresponds to the length of the segments in the 100-car baseline dataset. The visual comparison shows that the off-road glance duration distribution matches well with the data reported in the three related studies. However, the on-road glances show some differences between our data and the data reported in the 100-Car study and the study of \citet{Morando.2019}. Whereas the mode is similar for all three datasets, the on-road glances in our study tend to be shorter compared to the other two studies. The potential reasons for this are manifold. For example, \citet{Morando.2019} only consider driving segments of very controlled driving by excluding curved driving, lane changes, and driving segments with a vehicle speed under 60\,km/h. In these rather calm driving situations, drivers need to switch less often between the road and off-road regions such as mirrors or side windows resulting in longer continuous on-road glances. The differences with regard to the 100-Car study could be due to the fact that the data is now almost 20 years old and only covers manual driving. The technology of the vehicles at that time, and in particular that of the infotainment and assistance systems, was fundamentally different from that in today's vehicles. However, considering the differences in the data collection, the comparison suggests that our data collection and processing pipeline produces representative data.}

Figure~\ref{subfig:glance_duration_on_road} indicates that during touchscreen] interactions, drivers need to distribute their visual attention between the road and the center stack resulting to shorter on-road glances. On the other hand, center stack glances during secondary task engagements tend to be longer than general center stack glances (Figure~\ref{subfig:glance_duration_hu}). 
Through Figure~\ref{subfig:num_interactions}, we see that roughly 25\,\% of all sequences consist of only a single interaction. This results in many short secondary task engagements that only consist of a single glance toward the center stack (Figure~\ref{subfig:num_glances_hu}). These short engagements are part of real-world user behavior. However, they are often not represented in laboratory studies where only a few predefined tasks are evaluated. We argue that it is still relevant to analyze these short engagements and therefore decide to consider them.
For the long glance classification task, we balanced the dataset by applying random undersampling. The resulting dataset consists of 4,816 sequences for each class.

\subsection{Experimental Results}

We evaluate the regression models using a repeated 10-fold cross-validation \citep{Kohavi.1995} and the classification models using a stratified 10-fold cross-validation. The results are given in Table~\ref{tab:resultComparison}. The models were fitted on the full set of input features given in Table~\ref{tab:InputFeatures}.

\begin{table}[align=\centering]
\small
	\caption{Comparison of the different models.}
	\label{tab:resultComparison}
		\begin{tabular}{lrrrr}
			\toprule
			 & \multicolumn{2}{c}{Long Glance Prediction} & \multicolumn{2}{c}{Total Glance Duration Prediction}\\
			\cmidrule(lr){2-3}
			\cmidrule(lr){1-5}
			Model 					& Accuracy & Standard Deviation		& Mean Absolute Error 	& Standard Deviation\\ 
			\toprule
			Baseline		            & 50.09\,\%             & 1.63\,\%	& 4378\,ms 	& 177\,ms\\
			Logistic/Linear Regression	& 61.93\,\% 	        & 1.69\,\%	& 3778\,ms 	& 383\,ms\\
			\textbf{Random Forest} 	    & \textbf{67.53\,\%}	& \textbf{1.38\,\%}	    & \textbf{2437\,ms} & \textbf{112\,ms}\\
			XGBoost 	                & 67.22\,\%	&  1.85\,\%    & 2385\,ms	 & 117\,ms\\
			\ac{FNN}        & 65.90\,\% 	&  2.02\,\%    & 2443\,ms	 & 109\,ms\\
			\bottomrule
		\end{tabular}
\end{table}

The machine learning-based approaches outperform the Baseline approach and the Logistic and Linear Regression approaches in both tasks. The differences in the prediction accuracy support our assumption that neither of the problems at hand can be considered a linear problem and that interaction effects between different features exist. The machine learning models provide similar results. However, the Random Forest approaches offer two desirable properties making them in particular suitable for the use case at hand. First, the TreeSHAP \citep{Lundberg.2020} algorithm allows efficient computation of exact SHAP values for Random Forest models. Second, Random Forests can be run in parallel, making them suitable for future use cases when they are deployed on data of a whole production fleet. Thus, we choose the Random Forest models for the following explanation generation.

\subsection{Explainable Predictions}
While the above-presented results provide a good measure of prediction accuracy, they are of limited value when it comes to understanding human behavior. To truly support researchers and practitioners in the design process to foster a deeper understanding of drivers' visual attention allocation, it needs more than just predicting whether a new user flow might cause too much distraction~\citep{Ebel.Narrowing.2021}. For this reason, we employ \ac{SHAP}. \ac{SHAP} values represent the features' contribution to the model's output, providing a local explanation for each input sample. By combining many local explanations, one can represent global structures producing detailed insights into model behavior~\citep{Lundberg.2020}.

\subsubsection{Local Explanations}\label{ch:localExplanations}
\begin{figure}[pos=htpb!, align=\centering, width=\linewidth]
		\centering
		\subfloat[Long Glance Prediction Sample. ]{\includegraphics[width=\linewidth]{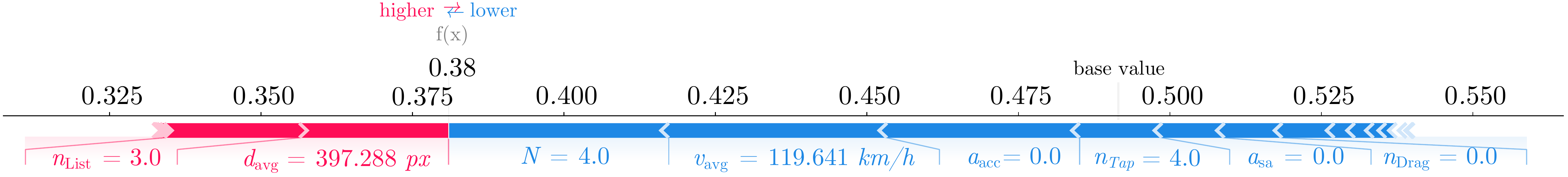}\label{fig:ForcePlotClassification}}
		\newline        \subfloat[Total Glance Duration Prediction Sample.]{\includegraphics[width=\linewidth]{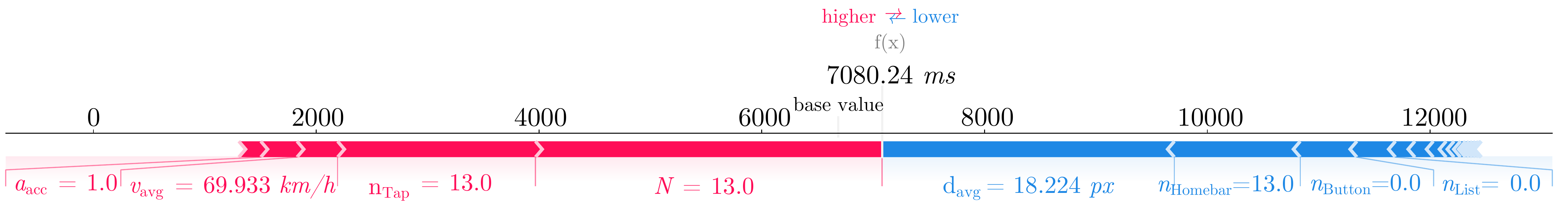}\label{fig:ForcePlotRegression}}
		\caption{Local explanations visualized as force plots.}
		\label{fig:ForcePlots} 
	\end{figure}

Figure~\ref{fig:ForcePlots} displays the explanations for one long glance prediction and one \ac{TGD} prediction. These force plots represent a particular model output as a cumulative effect of feature contributions (i.e. SHAP values). The length of each bar indicates how much the associated feature value pushes the model output from the base value toward higher values (red, to the right) or lower values (blue, to the left). The base value is computed as the average model output over the training dataset. The features in each group are sorted based on the magnitude of their impact and only the most influential features are displayed. The feature values are shown below the bars. For the long glance prediction, feature contributions are displayed as probabilities. For the \ac{TGD} prediction, they are shown in milliseconds.

Figure~\ref{fig:ForcePlotClassification} visualizes the explanation of a secondary task engagement for which the model outputs a long glance probability of $0.38 = 38\,\%$.
The long glance probability is pushed to the left because the driver only performed $4$ interactions ($N=4$) and drove at a speed of \fix{$v_{\text{avg}} = 119.641\,km/h$} while the ACC was deactivated ($a_{\text{acc}} = 0$). On the other hand, the prediction is pushed to the right because the interactions were quite distributed over the screen ($d_{\text{avg}} = 397.288\,px$) and three of them were list interactions ($n_{\text{List}} = 3$).

Another secondary task engagement is explained in Figure~\ref{fig:ForcePlotRegression}. Here, the \ac{TGD} prediction of roughly $7\,s$ is close to the base value because the positive and negative feature contributions balance each other out. During this secondary task engagement, the driver performed 13 touch interactions ($N$ = $n_{\text{Tap}} = 13$) while driving with an active ACC ($a_{\text{acc}} = 1$) at a speed of 70\,km/h. If the model would only access this information, it would predict a TGD of roughly 13 seconds However, as all interactions were very close to each other ($d_{\text{avg}} = 18.224\,px$) and were all performed on the homebar ($n_{\text{Homebar}} = 13$ without any list or button interaction interfering ($n_{\text{List}} = 0$, $n_{\text{Button}} = 0$, the final model output is only slightly higher than the average TGD prediction.

These local explanations show that not all features are always relevant. Predictions for secondary task engagements can be driven by only a few dominant features. The presented explanations enable designers and researchers to quickly identify the main forces behind individual predictions. It also allows them to play around with artificial input samples and observe how certain changes in the design of a user flow or the driving situation impact the model's output.

\subsubsection{Global Explanations}

\begin{figure}[pos=htpb!, align=\centering, width=\linewidth]
	\subfloat[Long Glance Prediction\label{subfig:BeeswarmClassification}]{%
		\includegraphics[width=0.49\linewidth]{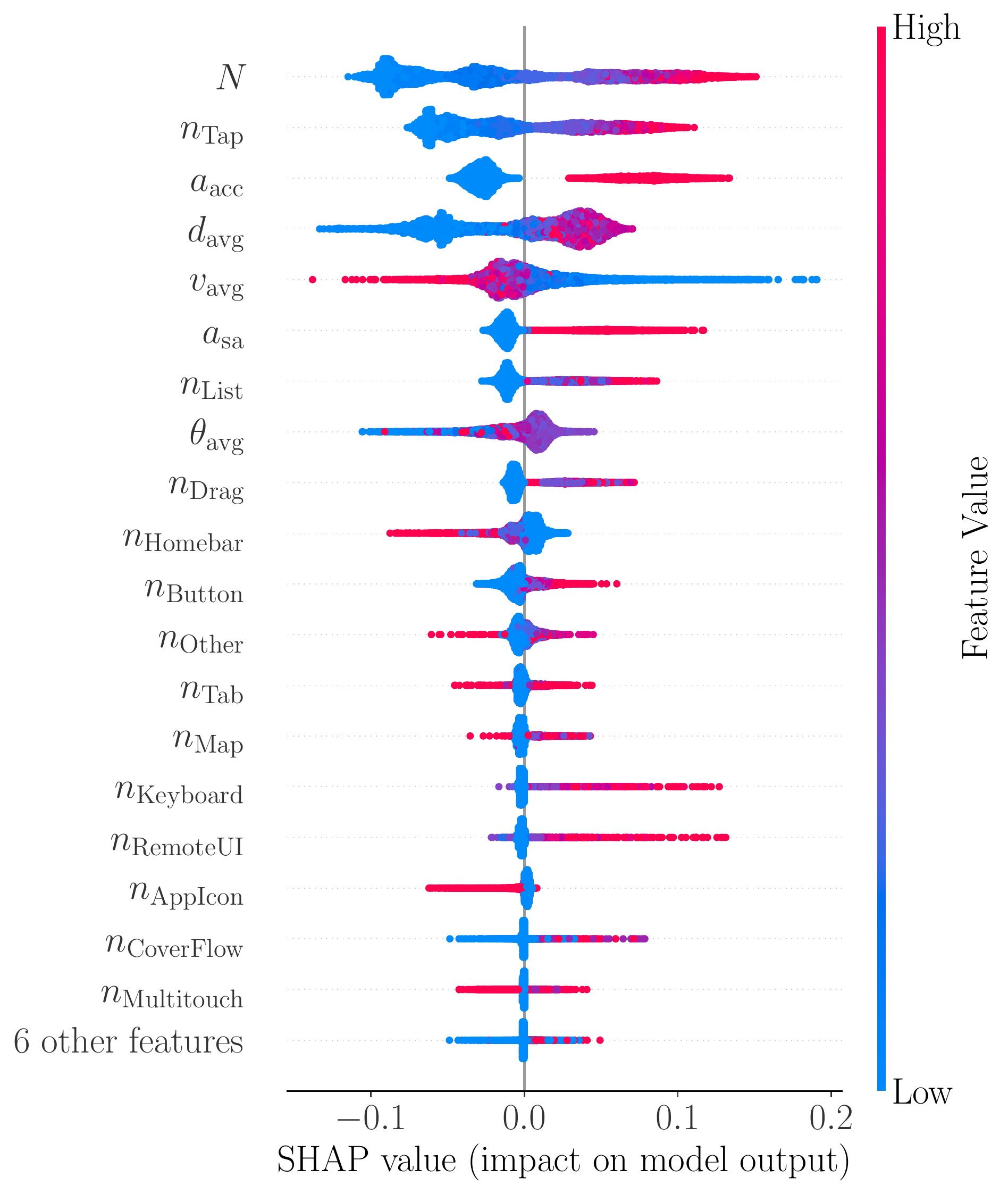}
	}
	\hfill
	\subfloat[Total Glance Duration Prediction\label{subfig:BeeswarmRegression}]{%
		\includegraphics[width=0.49\linewidth]{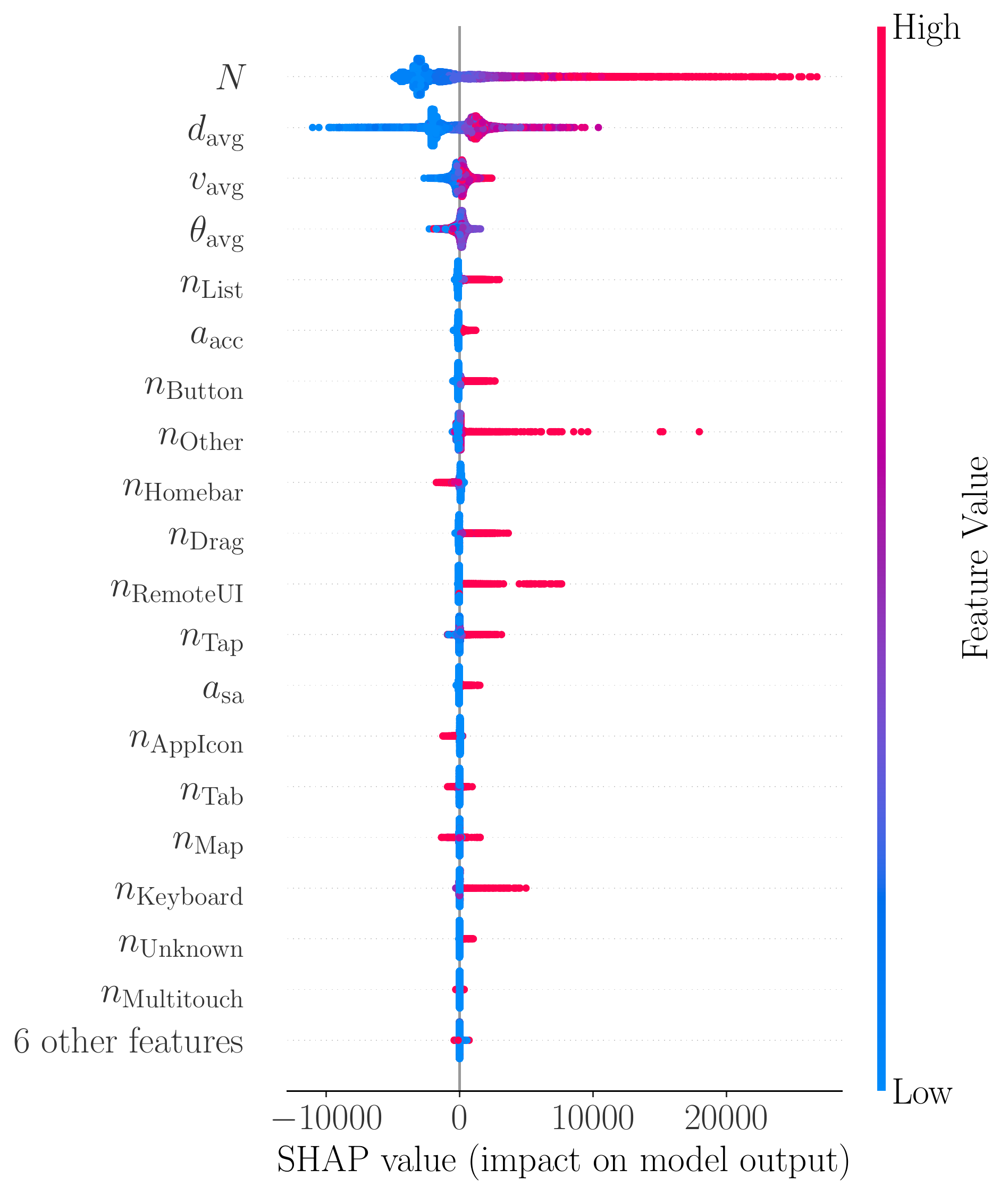}
	}
	\caption{Explanation summary visualized as a set of beeswarm plots. Each beeswarm plot represents the distribution of SHAP values for one feature.}
	\label{fig:Beeswarm}
\end{figure}

To understand how the features affect the model's output on a global scale, we combine all local explanations of the dataset. Figure~\ref{fig:Beeswarm} shows the distribution of SHAP values (i.e., the impact of each feature on a specific prediction as seen in \ref{ch:localExplanations}) as a set of beeswarm plots. Each dot in a row corresponds to an individual secondary task engagement. The position on the x-axis represents the effect of the respective feature on the model's output. In Figure~\ref{subfig:BeeswarmClassification}, the SHAP values are in probability space, and in Figure~\ref{subfig:BeeswarmRegression} they represent the impact in milliseconds. The color indicates the feature value (red is high, blue is low). The features are sorted by their global importance and only the 19 most important features are displayed individually.

The most important features of the long glance prediction model (Figure~\ref{subfig:BeeswarmClassification}), are the number of interactions $N$, the average distance between the interactions $d_{\text{avg}}$, and the number of tap gestures $n_{\text{Tap}}$. The more touchscreen interactions a driver performs and the larger the distance between them, the higher the output probability that one of the associated glances is longer than 2 seconds. Figure~\ref{subfig:BeeswarmClassification} also reveals that both, the activation of \ac{ACC} $a_{\text{acc}}$ and \ac{SA} $a_{\text{sa}}$, increase the long glance probability. Whereas the impact of a deactivated assistance system (blue) is small for all samples, the impact varies if the assistance systems are active. The horizontal spread suggests that the impact of assisted driving on visual attention allocation is situation-specific and depends on further factors like the driving situation and interaction patterns. 
The distribution that describes the impact of the vehicle speed $v_{\text{avg}}$ is heavily tailed. For most secondary task engagements at medium speed, the effect is negative but rather small. High speed values reduce the predicted long glance probability and low speed values increase it, respectively. This indicates drivers' self-regulative behavior.

The number of list interactions $n_{\text{List}}$ is the most important feature associated with a specific UI element followed by the number of interactions with the homebar $n_{\text{Homebar}}$. Through Figure~\ref{subfig:BeeswarmClassification}, we see that their impact is opposite to each other. Whereas the long glance probability increases with an increasing number of list interactions, it decreases for an increasing number of homebar interactions. This suggests that list interactions tend to be more distracting than interactions on the static homebar. The impact of interactions with Android Auto or Apple Car Play $n_{\text{RemoteUI}}$ is similar to the impact of list interactions.
In general, we can observe that most of the SHAP value distributions associated with a specific class of UI elements are centered around zero with long tails to one or both sides.
This is because most of the elements occur in only a small portion of secondary task engagements. Whereas this leads to a relatively low global importance, these features still have a large impact on specific predictions.

For the \ac{TGD} prediction model (Figure~\ref{subfig:BeeswarmRegression}), $N$ and  $d_{\text{avg}}$ are also the two most important features. Their distributions also show similarities to the distributions observed in the long glance prediction task. However, the impact of the vehicle speed $v_{\text{avg}}$ is inverse compared to the long glance prediction task. High speed values increase the \ac{TGD} prediction and low values decrease the prediction. Both findings together could be an indication that drivers reduce their single glance duration at higher speeds, which in turn results in longer TGDs because more individual glances are required to complete the same task. 

\begin{figure}[pos=htpb!, align=\centering]
	\subfloat[ACC\label{subfig:scatter_plot_classification_acc}]{%
		\includegraphics[width=0.49\linewidth]{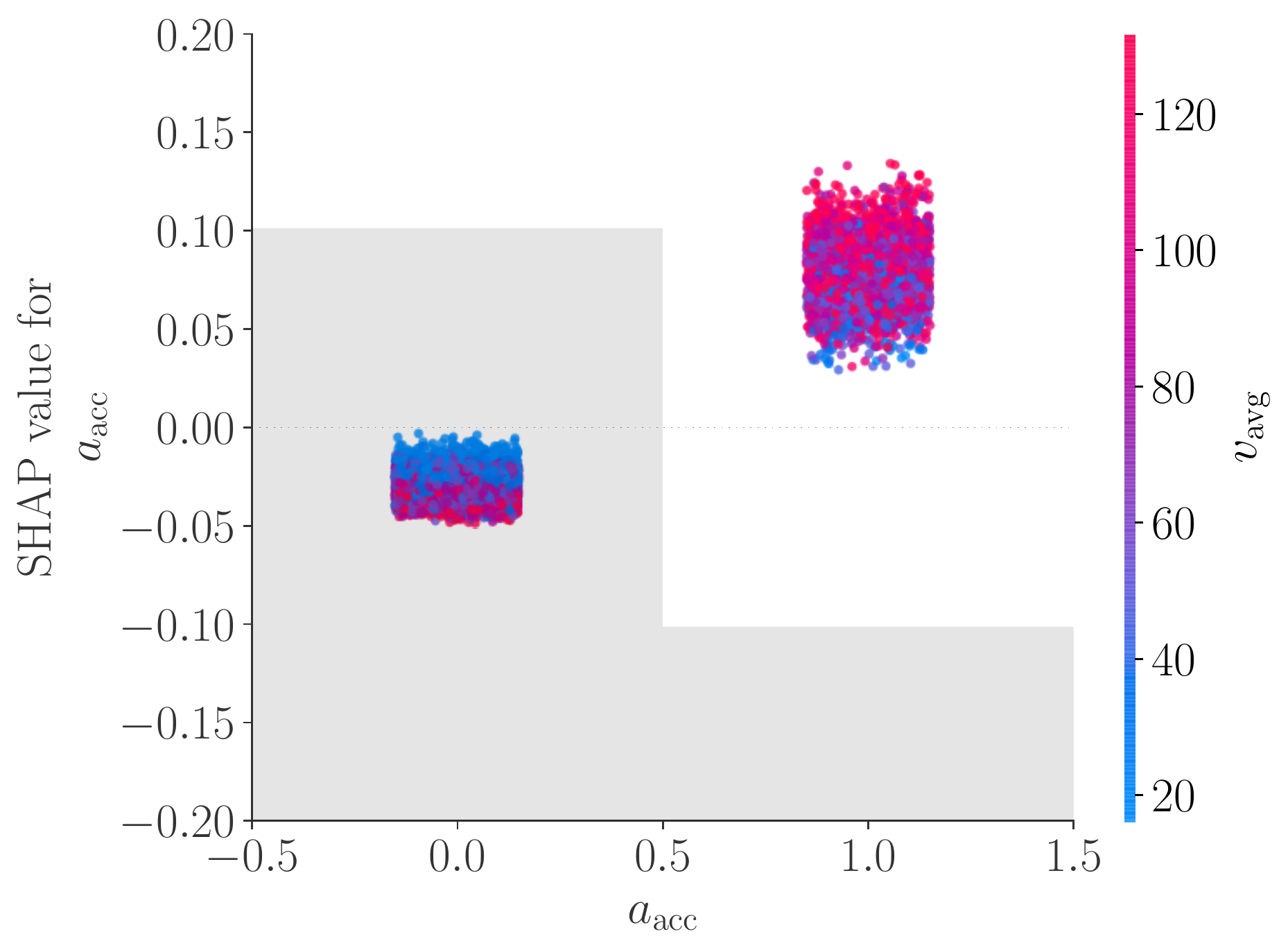}
	}
	\hfill
	\subfloat[Vehicle Speed\label{subfig:scatter_plot_classification_speed}]{%
		\includegraphics[width=0.49\linewidth]{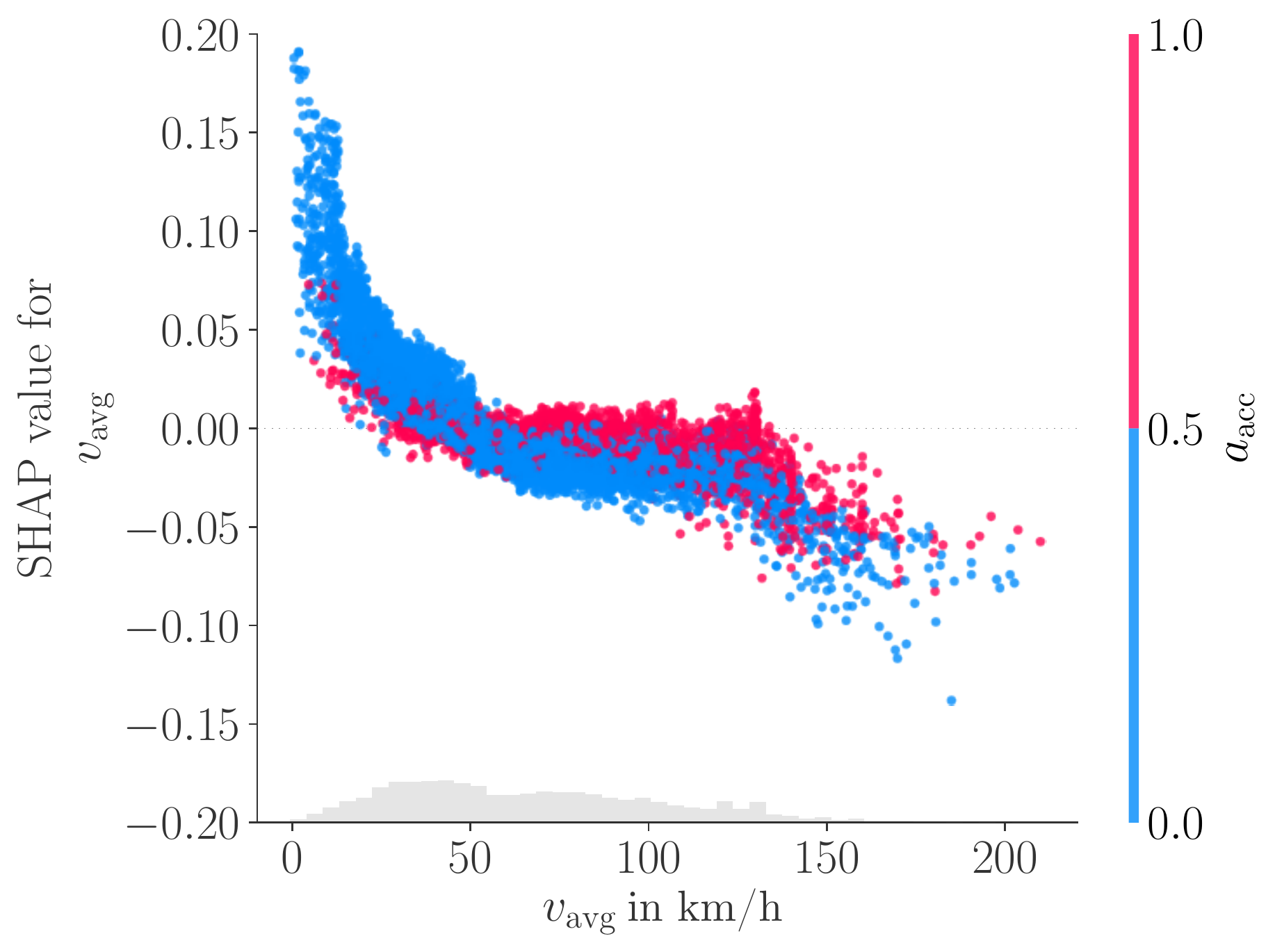}
	}
	\hfill
	\subfloat[N\label{subfig:scatter_plot_classification_N}]{%
		\includegraphics[width=0.49\linewidth]{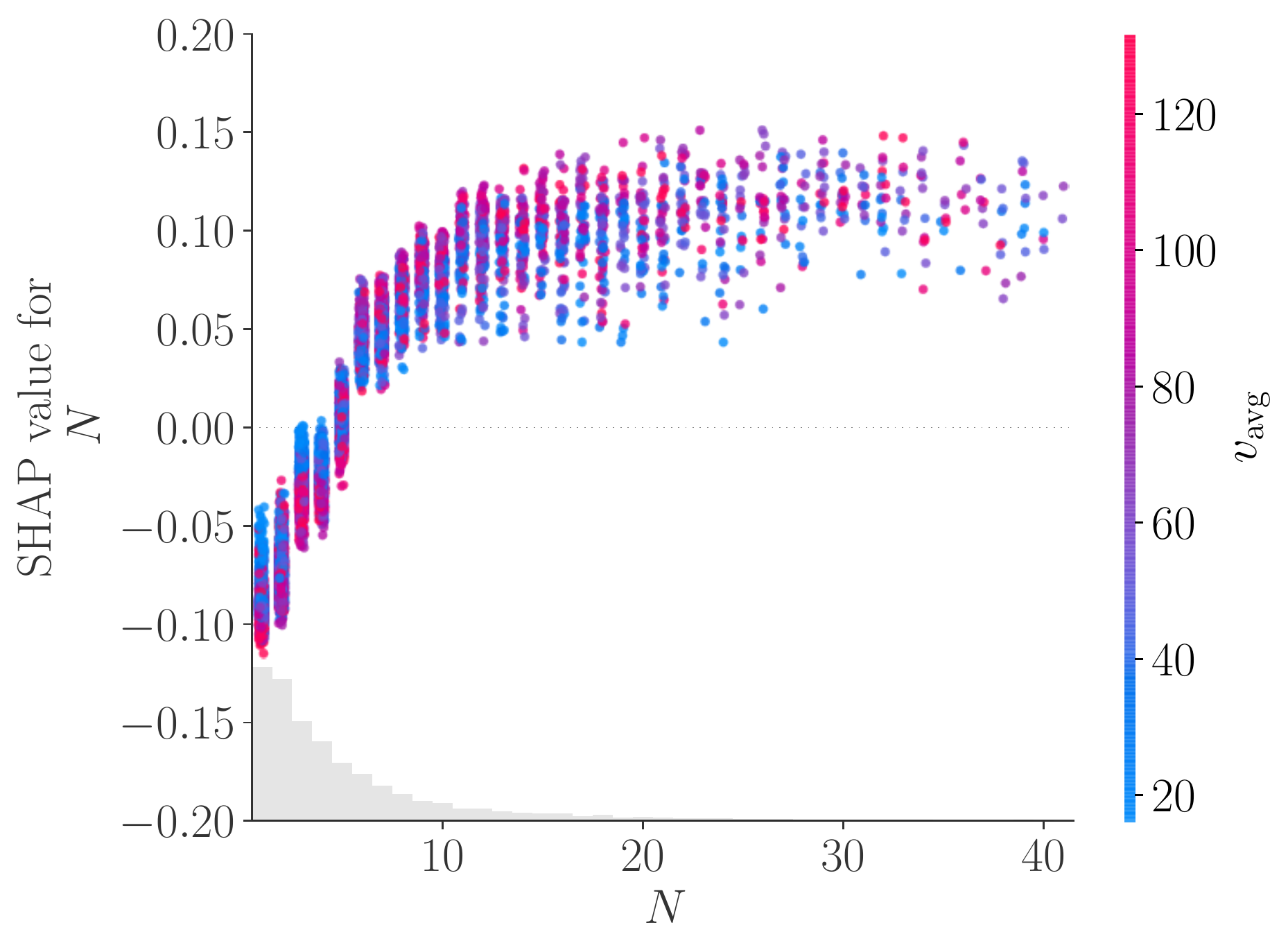}
	}	
	\hfill
	\subfloat[Touch Distance\label{subfig:scatter_plot_classification_avg_distance}]{%
		\includegraphics[width=0.49\linewidth]{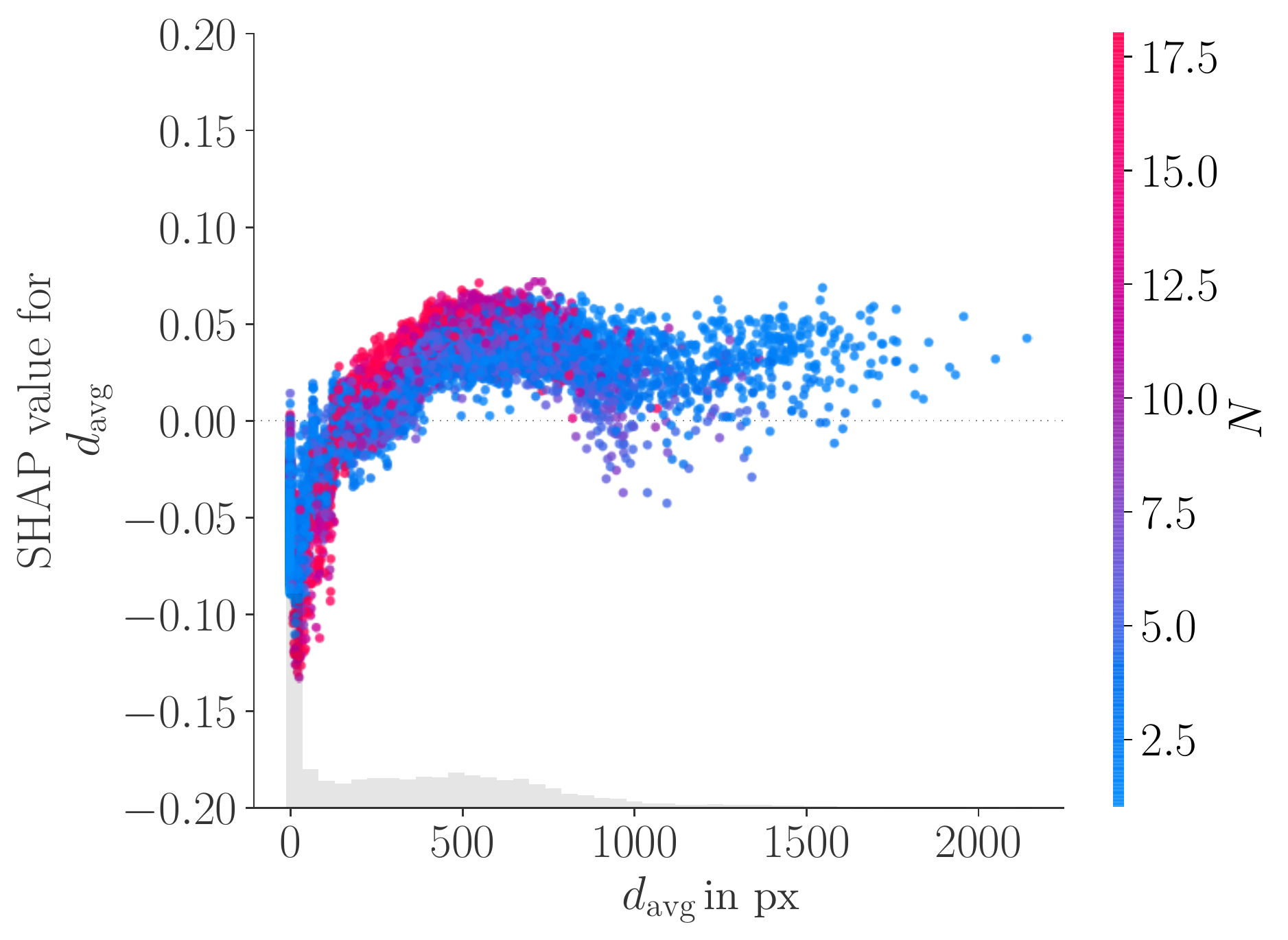}
	}
	\hfill
	\subfloat[Homebar\label{subfig:scatter_plot_classification_Homebar}]{%
		\includegraphics[width=0.49\linewidth]{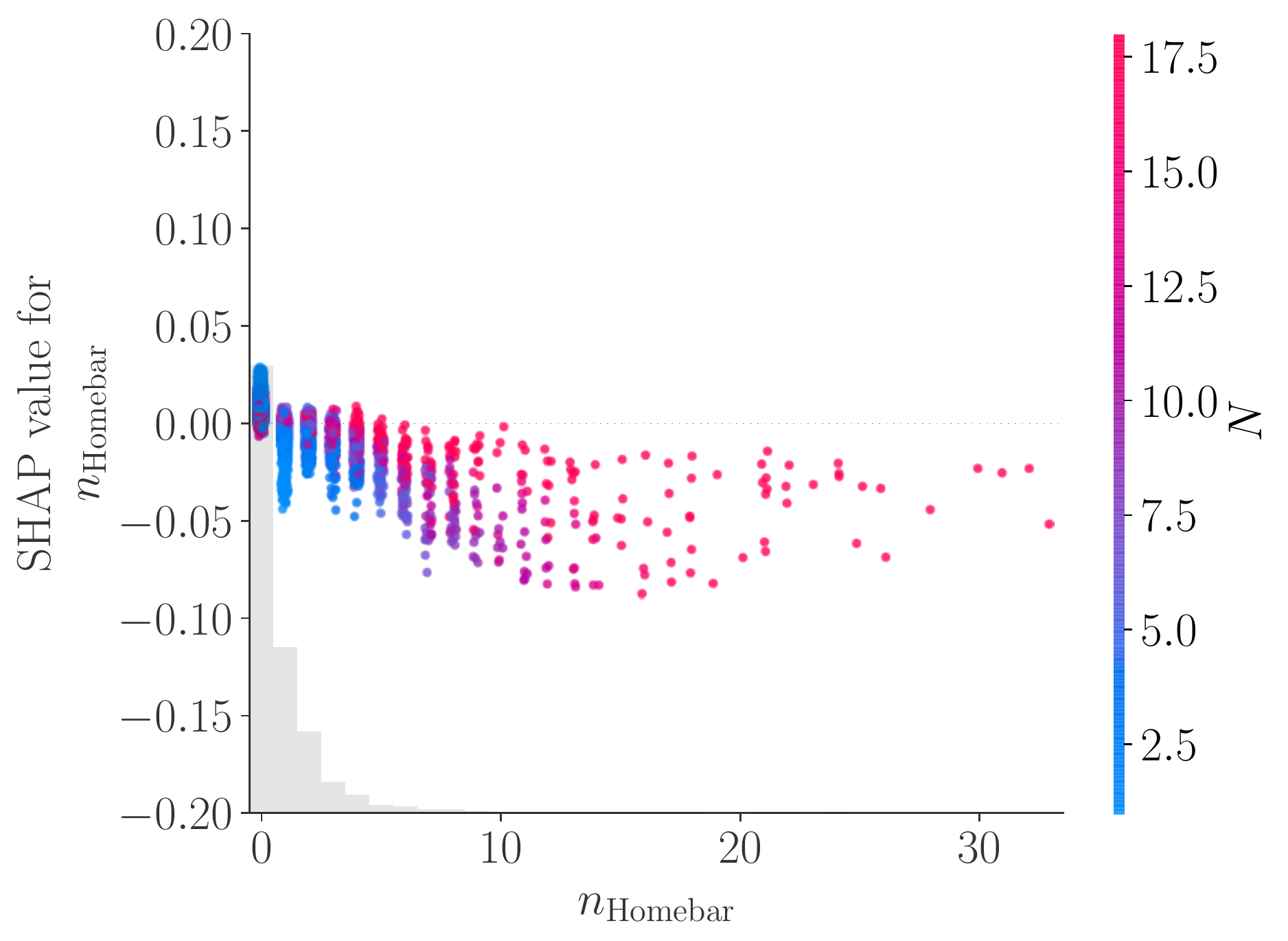}
	}	
	\hfill
	\subfloat[List\label{subfig:scatter_plot_classification_List}]{%
		\includegraphics[width=0.49\linewidth]{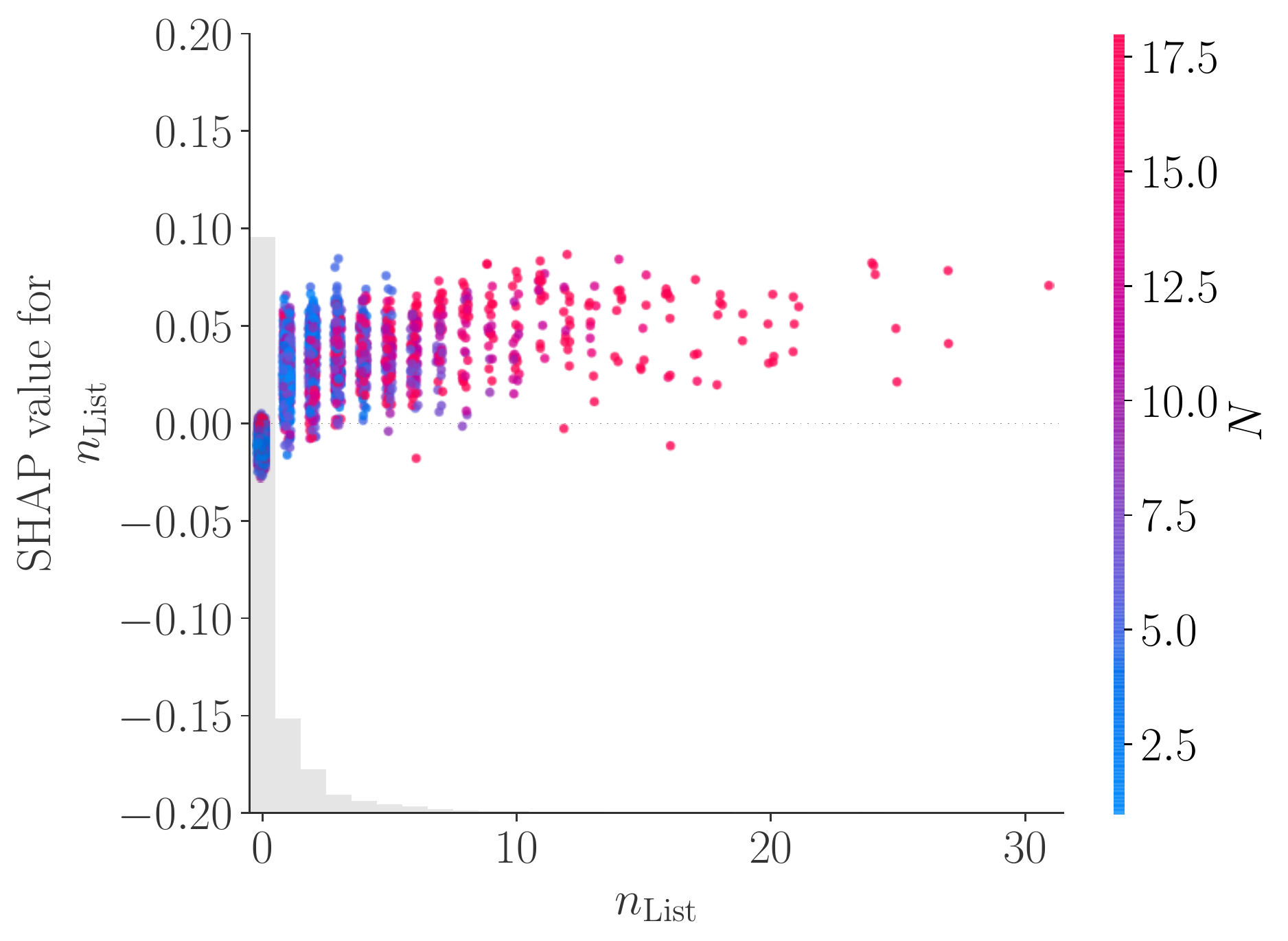}
	}
	\caption{Feature dependence plots for the long glance classification model.}
	\label{fig:ScatterLongGlance}
\end{figure}

Further, we can see that there are almost no negative contributions associated with UI interaction features. This is due to the fact that the TGD task is cumulative, and every interaction inevitably implies a certain amount of visual attention. However, homebar interactions $n_{\text{Homebar}}$, can negatively affect the model output. In line with the observations made for the long glance prediction, list interactions $n_{\text{List}}$, map interactions $n_{\text{Map}}$ and interactions with Android Auto and Apple CarPlay $n_{\text{RemoteUI}}$ can be associated with an increased visual demand prediction.
A comparison between Figure~\ref{subfig:BeeswarmClassification} and Figure~\ref{subfig:BeeswarmRegression} also reveals that the TGD is not as dependent on the status of the driver assistance systems as the long glance probability.

\begin{figure}[pos=htpb!, align=\centering]
	\subfloat[ACC\label{subfig:scatter_plot_regression_acc}]{%
		\includegraphics[width=0.49\linewidth]{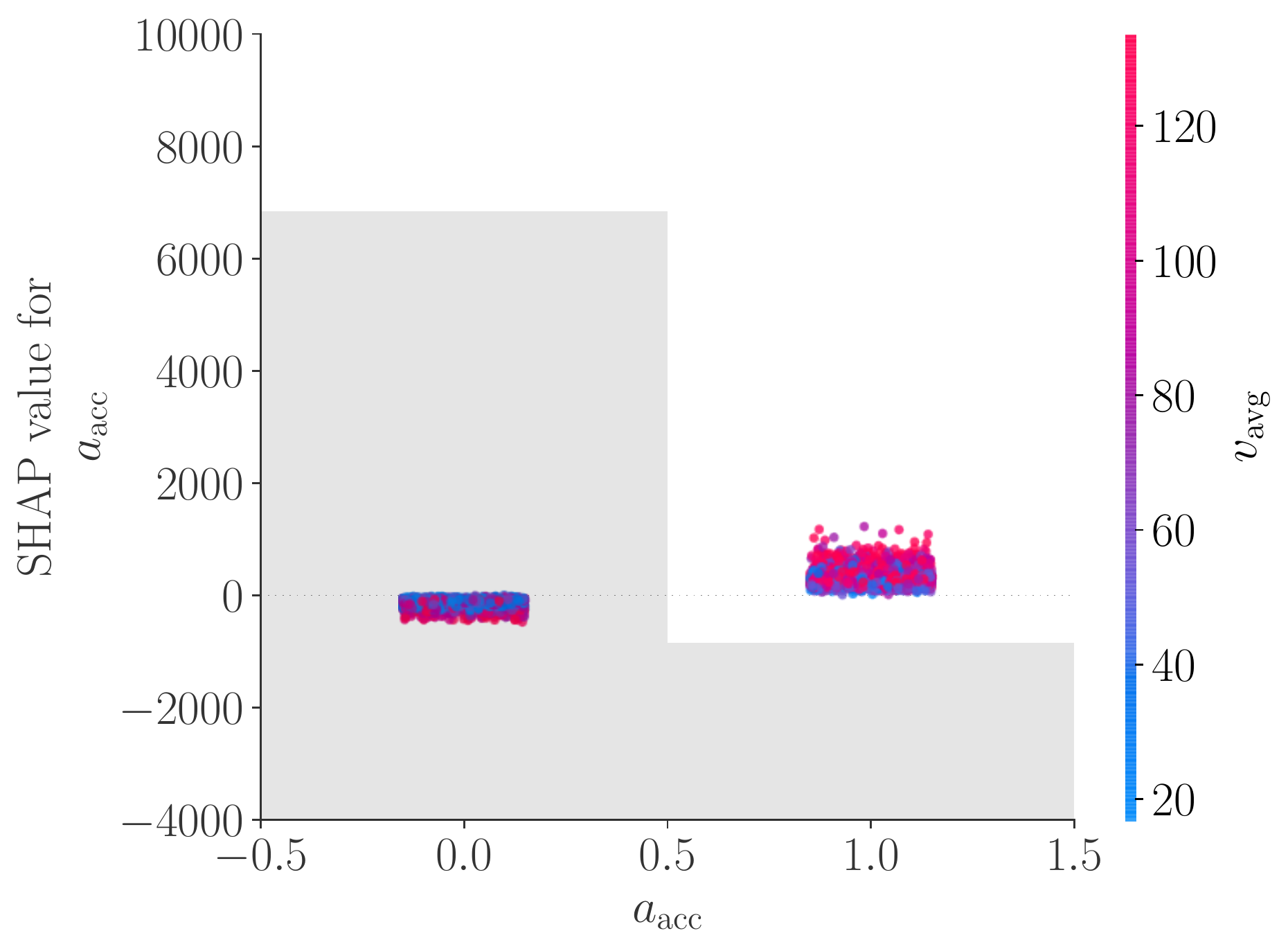}
	}
	\hfill
	\subfloat[Vehicle Speed\label{subfig:scatter_plot_regression_speed}]{%
		\includegraphics[width=0.49\linewidth]{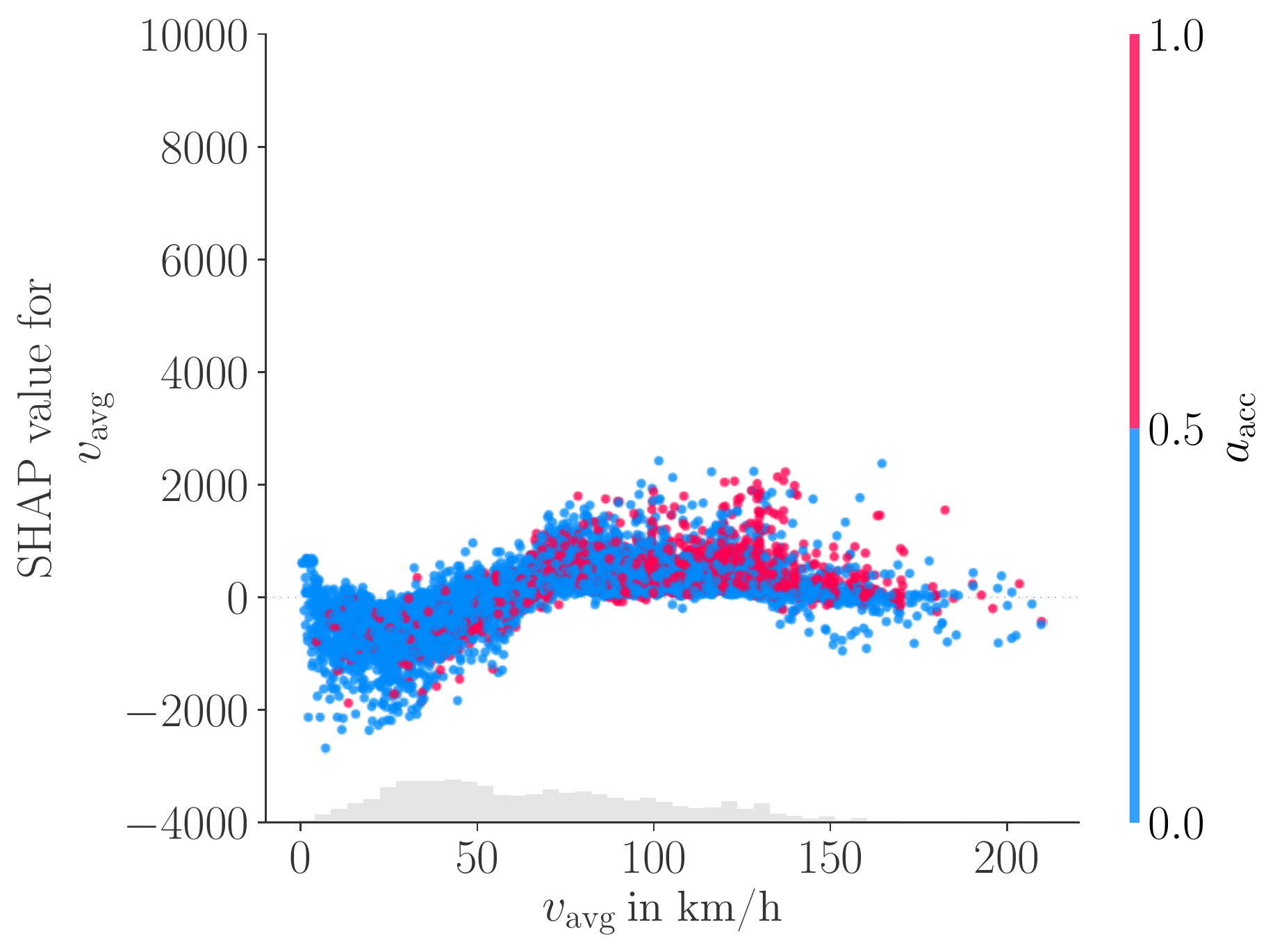}
	}
	\hfill
	\subfloat[N\label{subfig:scatter_plot_regression_N}]{%
		\includegraphics[width=0.49\linewidth]{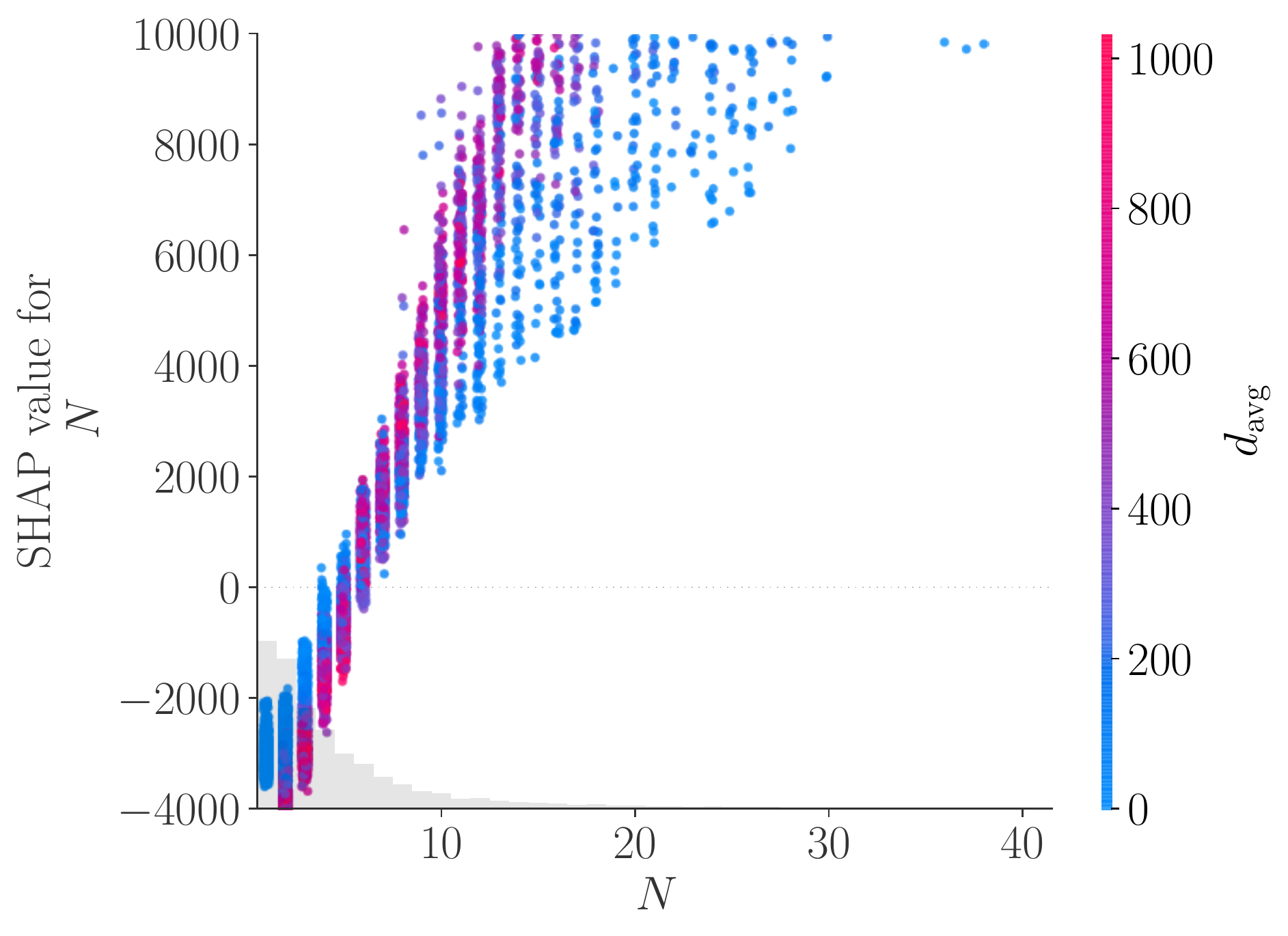}
	}	
	\hfill
	\subfloat[Touch Distance\label{subfig:scatter_plot_regression_avg_distance}]{%
		\includegraphics[width=0.49\linewidth]{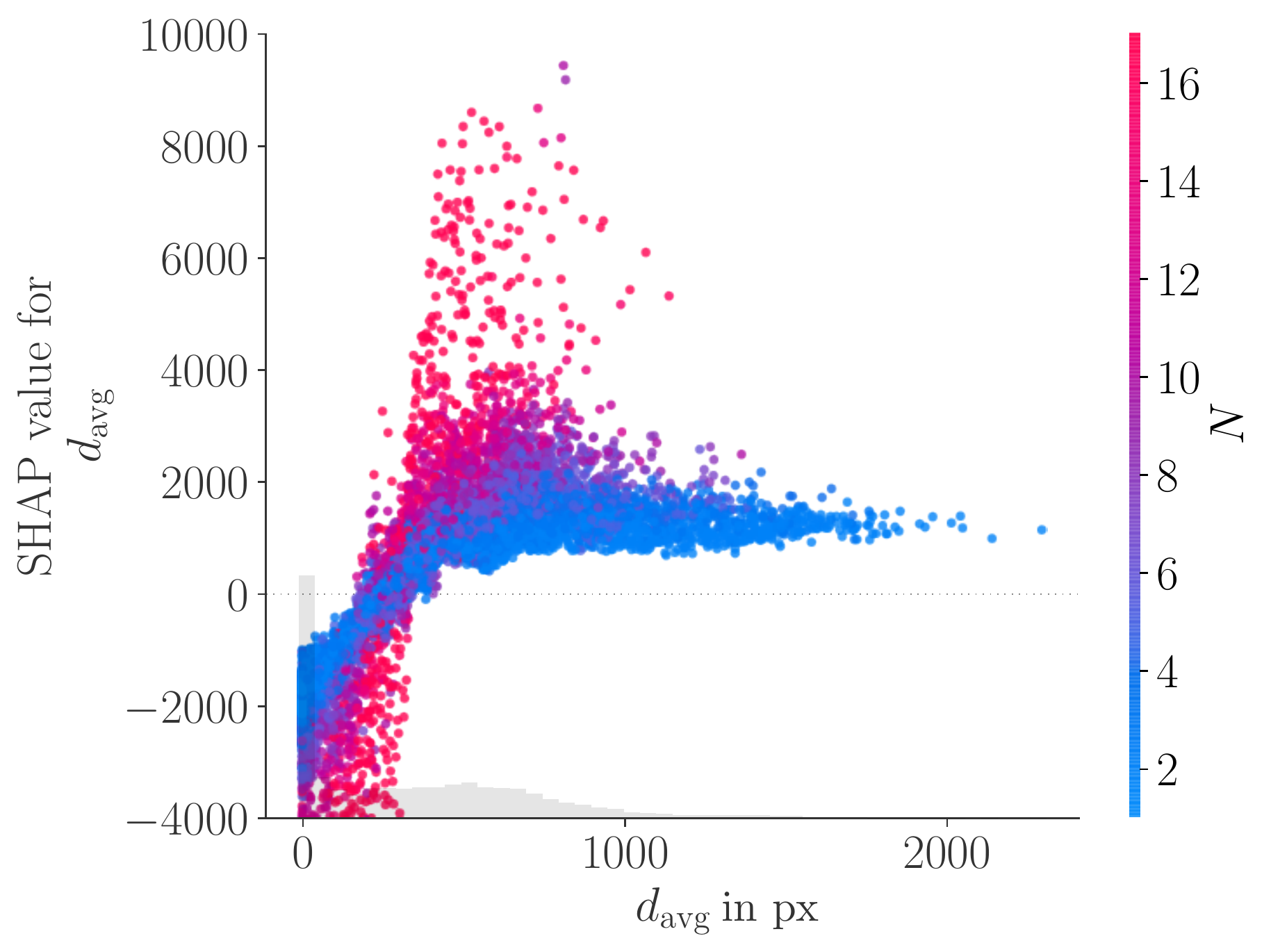}
	}
	\hfill
	\subfloat[Homebar\label{subfig:scatter_plot_regression_Homebar}]{%
		\includegraphics[width=0.49\linewidth]{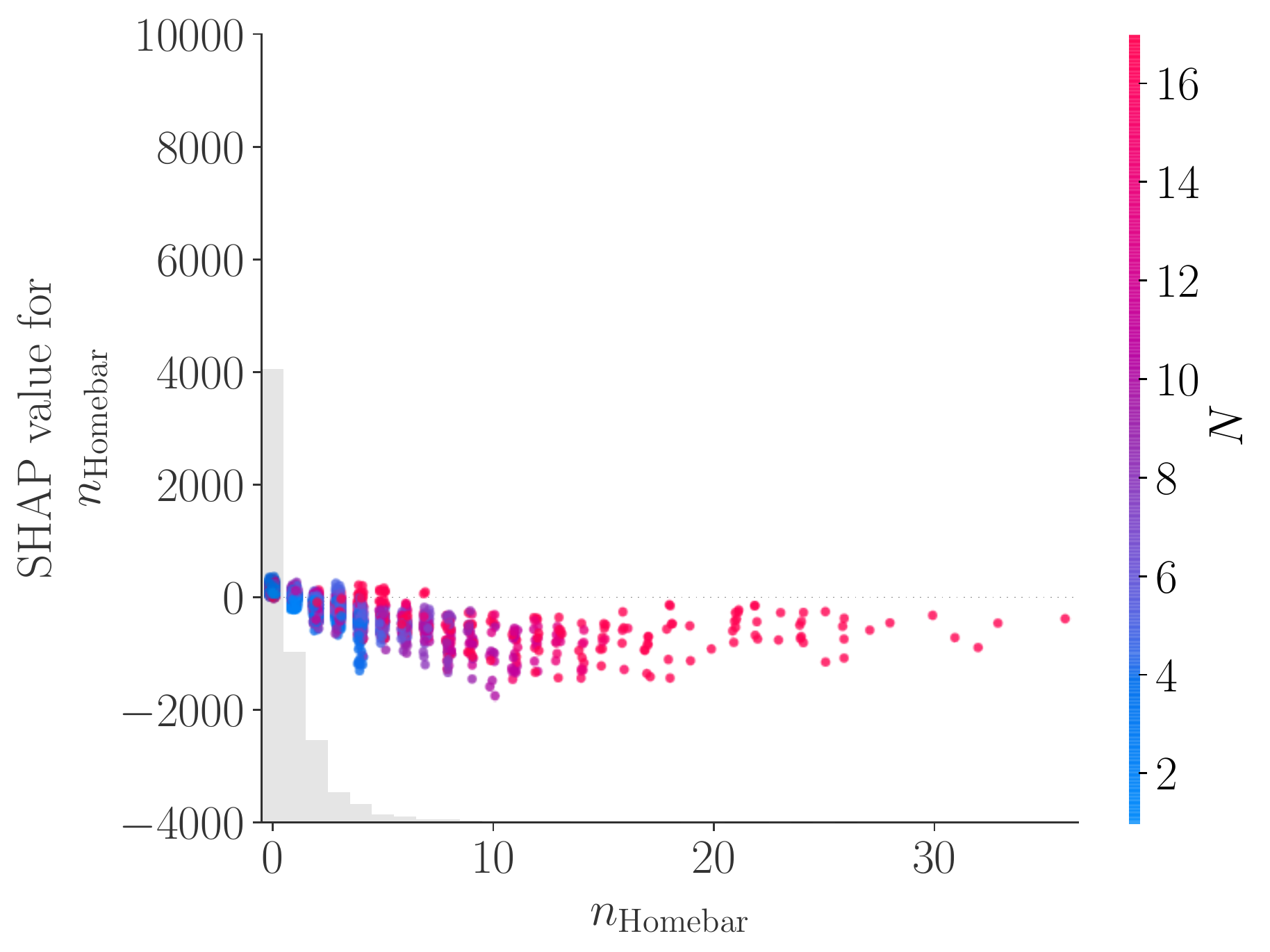}
	}	
	\hfill
	\subfloat[List\label{subfig:scatter_plot_regression_List}]{%
		\includegraphics[width=0.49\linewidth]{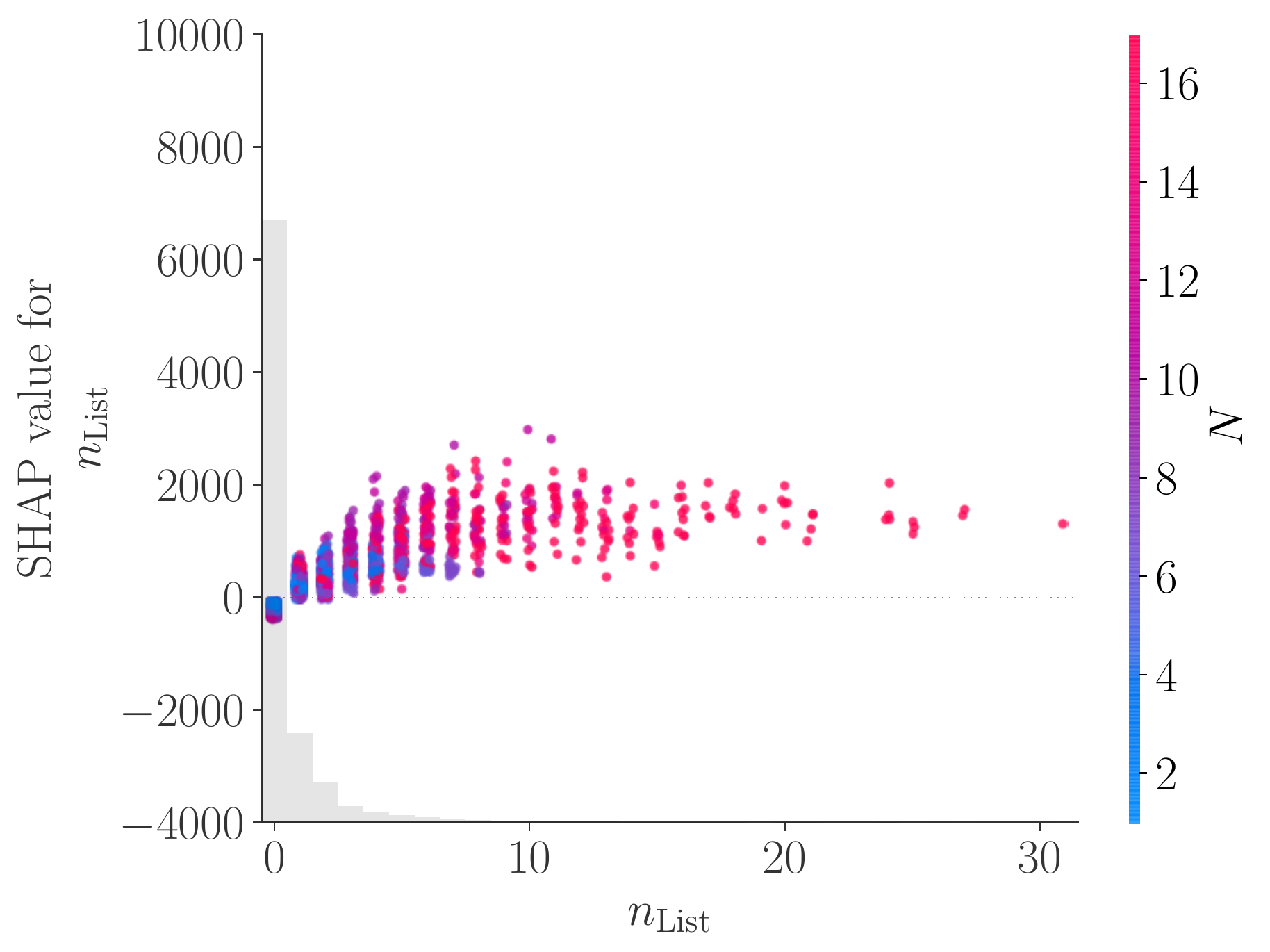}
	}
	\caption{Feature dependence plots for the TGD model.}
	\label{fig:ScatterDuration}
\end{figure}

To understand the effect of a single feature on the model's output in more detail, we plot the SHAP values (y-axis) against the corresponding feature values (x-axis). Every secondary task engagement in our dataset is represented as a dot (see Figure~\ref{fig:ScatterLongGlance} and Figure~\ref{fig:ScatterDuration}). Vertical dispersion at a single value on the x-axis shows that there are non-linear dependencies between the displayed feature and other features. To highlight the interaction between features, each dot is colored by the value of the feature that shows the strongest interaction. The histogram at the bottom of the plots shows the distribution of datapoints.
Figure~\ref{subfig:scatter_plot_classification_acc} suggests that the use of \ac{ACC} leads to an increased long glance probability prediction. The interaction with the vehicle speed shows that the effect tends to increase with increasing vehicle speed. On the other hand, the data shown in Figure~\ref{subfig:scatter_plot_classification_speed} indicates that drivers tend to increase their single glance durations at lower speeds (below 50\,km/h) and decrease them at higher speeds (above 125\,km/h). However, in between those values, the speed has almost no influence on the model output. This suggests that drivers self-regulate their visual attention allocation based on what they consider an appropriate speed. Additionally, the interaction with the ACC status shows that the impact of the speed on the model output decreases when ACC is active. The interaction effect with $a_{\text{acc}}$ partially explains the variance (vertical diversion) in the effect of the vehicle speed. However, various factors like road type or speed limit that may also influence how the vehicle speed affects drivers' visual attention allocation are not considered in the presented models.

\fix{Figure~\ref{subfig:scatter_plot_classification_N} indicates that the number of interactions is positively correlated with the drivers' probability to perform a long glance. On the other hand, Figure~\ref{subfig:scatter_plot_classification_avg_distance} suggests that as soon as the distance between the touch interactions exceeds a certain threshold (roughly 200\,px), the effect on the long glance probability remains constant.
Whereas homebar interactions decrease the probability of the model predicting a long glance (Figure~\ref{subfig:scatter_plot_classification_Homebar}),} list interactions (Figure~\ref{subfig:scatter_plot_classification_List}) push the model toward predicting a long glance. The interaction effect with the number of interactions additionally indicates that the impact of both elements becomes larger the higher their proportion within a sequence is.

\fix{Figure~\ref{fig:ScatterDuration} visualizes how the different features affect the TGD prediction. While the number of interactions $N$ (Figure~\ref{subfig:scatter_plot_regression_N}) is the dominant feature it is also the feature with the highest interaction effect on all other features.
Compared to Figure~\ref{fig:ScatterLongGlance} and in line with the observations we made in Figure~\ref{fig:Beeswarm}, we see that the ACC status $a_{\text{acc}}$ (Figure~\ref{subfig:scatter_plot_regression_acc}) and the vehicle speed $v_{\text{avg}}$ (Figure~\ref{subfig:scatter_plot_regression_speed}) do not influence the TGD prediction as much as they influence the long glance prediction. This applies in particular to secondary task engagements with few interactions.
The impact of list interactions $n_{\text{List}}$ and homebar interactions $n_{\text{Homebar}}$ on the TGD, however, is similar to the impact those interactions have on the long glance probability (Figure~\ref{subfig:scatter_plot_regression_Homebar} and Figure~\ref{subfig:scatter_plot_regression_List}). This also applies to the influence of the average touch distance $d_{\text{avg}}$.} An increase in touch distance leads to an increase in TGD and long glance probability until a certain threshold is reached. However, the interaction effect with $N$ is higher for the TGD prediction model. Another interesting aspect that might need further exploration is the location of the x-intercept. This point describes the touch distance at which the feature's impact turns from decreasing to increasing the visual demand prediction (Figure~\ref{subfig:scatter_plot_regression_avg_distance}).

\section{Discussion}

\fix{The presented approach enables users to evaluate the visual demand of early-stage prototypes. In the following, we put our results into perspective and show that the presented approach is more accurate than comparable methods. The predictions and explanations facilitate the generation of fast insights without requiring expensive and long-planned user studies. We illustrate this by assessing three exemplary research objectives covered in the literature. Finally, we address several limitations that apply to our approach.}

\subsection{\fix{Predicting the Visual Demand of In-Vehicle Touchscreen Interactions}}

\fix{Given the complexity of the modeling task, the presented results show how machine learning methods can be used to generate valuable insights into drivers' multitasking behavior by leveraging large naturalistic driving data. Compared to the approach of \citet{Kujala.2015}, who report critical differences between model predictions and observations, our approach is not only more accurate but also considers a more diverse set of UI elements.}

\fix{Our approach can predict the \ac{TGD} with a mean absolute error of roughly 2.4 seconds over a diverse range of interactions and driving scenarios. In comparison, \citet{Purucker.2017} report a mean error of 4\,s when averaged over all evaluated tasks. Furthermore, \citet{Purucker.2017} use a simple car following task at a constant speed for evaluation. Although these comparisons are useful to put the results into perspective, one needs to consider that the approaches highly differ in their environments and scenarios as described by \citet{Janssen.2020}.}

\subsection{\fix{Fast and Easily Accessible Insights based on Real-World Driving Data}}

Our approach has two main advantages over conventional user studies. First, the models allow making predictions for yet unseen secondary task engagements. Conventional studies can only be used to evaluate situations that were explicitly tested. Second, if the user interface undergoes disruptive changes (e.g., a completely new design or concept), the results of a user study are no longer valid and a new study needs to be conducted. Similarly, our computational models may also lose their capability to generalize. However, the advantage of the automated approach for data collection and modeling is that, as soon as a new version is deployed to test vehicles, data is collected and new models based on this new version can be fitted.
To demonstrate that our approach is a meaningful extension of traditional user research methods, we compare our results with those from conventional user studies.

\textbf{The Influence of Vehicle Speed on Drivers' Visual Attention Allocation.} 
\citet{Risteska.2021} found that an increase in speed reduces drivers' long off-path glances. They argue that drivers modulate their visual attention allocation based on driving demands. A similar finding is presented by \citet{Tivesten.2014}, who found a significant correlation between vehicle speed and off-road glance duration when drivers were engaged in a visual manual phone task. This is consistent with our results shown in Figure \ref{subfig:scatter_plot_classification_speed} and Figure \ref{subfig:BeeswarmClassification}. Our explanations do not only indicate that the long glance probability decreases with increasing speed but further suggest that this behavior might not be strictly proportional and is also affected by the status of driver assistance systems.
Our results further show that the predicted TGD increases with increasing speed (see Figure \ref{subfig:scatter_plot_regression_speed}). The combination of both findings provides a more comprehensive picture suggesting that drivers reduce their single glance duration at higher speeds, forcing them to look to the center stack touchscreen more often. This, in turn, leads to increased TGDs because certain aspects of human glance behavior like the time needed to locate an item are constant for each glance \citep{Large.2017}.

\textbf{The Influence of Driving Automation on Drivers' Visual Attention Allocation.} 
Assisted driving is associated with an increase in the mean and total glance duration during secondary task engagements \citep{Large.2017a, Carsten.2012, ebel.2022}. This is in line with our findings presented in Figure \ref{fig:Beeswarm}. In a driving simulator study, \citet{Carsten.2012} also found that the effect of lateral control (SA) on driver engagement is larger than the effect of longitudinal control (ACC). Based on our data, we cannot confirm this finding. The reasons for this can be manifold but may well be due to the difference between real data and simulation data. Our results further differ from those of \citet{Morando.2019}, who report no differences in the aggregate off-path glance duration distributions between manual and assisted driving. They only report an effect concerning the on-road glance distribution but state that their eye-tracker did not provide detailed information about the off-path \acp{AOI}. Since we can explicitly detect glances toward the center stack touchscreen and can distinguish them from general off-path glances, we argue that our results are superior.

\textbf{The Influence of Design Characteristics on Drivers' Visual Attention Allocation.}
There are not yet many approaches that have investigated the influence of design characteristics on visual demand in such detail (element type basis) as we show in our approach. \citet{Kujala.2015} found that the average distance between two consecutive touch interactions is a critical factor associated with long glances exceeding the limit considered safe. This is in line with our results presented in Figures~\ref{subfig:scatter_plot_classification_avg_distance}~and~\ref{subfig:scatter_plot_regression_avg_distance}. The explanations that our method provides could additionally serve as a first attempt to quantify the impact spatial separation of interaction elements has on visual demand while driving. Our approach also allows us to make detailed statements about the influence of individual elements. So far, only the task interaction times have been studied in the literature in a roughly similar level of detail \citep{Green.2015, Schneegass.2011}. We found that in particular interactions with maps, lists, and interactions within Apple CarPlay and Android Auto seem to be visually demanding. Interactions on the static homebar, with app icons, and general buttons, on the other hand, are less demanding.

\subsection{Benefits for the Design Process of IVISs and Implications on Distracted Driving Prevention}

To develop \acp{IVIS} that are safe to use, driver distraction evaluation needs to be an integral part already in the early design stages. However, driver distraction is a complex construct, and automotive UX experts need data-driven support to evaluate and compare design alternatives concerning their distraction potential~\citep{Ebel.Narrowing.2021}. Thus, our approach aims to inform the design process of \acp{IVIS} from the bottom up to develop solutions that are the least distracting and safe by design. We envision our method to be used to dynamically evaluate early-stage \ac{IVIS} designs. Users can assess hypothetical IVIS designs concerning their distraction potential in terms of visual demand. They can play around with artificial input samples to learn how changes in the user flow or driving scenario affect drivers' visual attention allocation. Our method then explains how each parameter contributes to the overall prediction. Thus, designers can better understand the effects of various UI elements, driving automation, and vehicle speed on driver distraction. This information can then be used to design \acp{IVIS} that are less distracting and reduce the risk of accidents. The improved accuracy over comparable approaches and the three application examples show that our approach can make a major contribution to better understanding the complex construct of driver distraction and drivers' visual attention allocation during secondary touchscreen tasks.

\subsection{Limitations and Future Work}

As we leverage already commercialized technologies of our research partner, we collected a large amount of behavioral data. We observed drivers' natural interaction behavior without explicitly telling them which touchscreen interactions to perform and therefore eliminate the so-called instruction effect \citep{carsten.2017}. While this approach has many advantages, especially over simulator and test track studies, several limitations apply. These limitations and their potential implications are discussed in the following.

Only company internal cars contributed to the data collection. Whereas they are used for a diverse range of testing procedures, they are also used for transfer and leisure rides of employees, for example over the weekend. We argue that, even if drivers follow a test protocol that aims to evaluate driving-related functions, the incentive to interact with the \acp{IVIS} does not deviate much from real-world behavior.
Furthermore, all drivers in this study need to be considered expert users. However, it is not yet entirely clear to what extent the gaze behavior of experts differs from regular users. Whereas \citet{wikman.1998} report that experienced drivers allocated their visual attention more adequately~\citep{wikman.1998}, \citet{naujoks.2016} show that experienced users of \acp{ADAS} tend to increase their secondary task engagements compared to novice users. However, a comparison with related approaches~\citep{Gaspar.2019, Morando.2019, noble.2021} shows high agreement in total and average glance behavior. \fix{Still, the restricted sample of drivers and the fact they were driving alone, need to be considered when interpreting the results.}

It is important to consider that the features used in this work do not capture all factors that influence drivers' visual attention allocation. In this study, we only consider the level of driving automation, vehicle speed, and the steering wheel angle to describe the driving situation. These features and their interactions provide valuable information (compare Figures~\ref{subfig:scatter_plot_classification_acc}, \ref{subfig:scatter_plot_classification_speed}, \ref{subfig:scatter_plot_regression_speed}, \ref{subfig:scatter_plot_regression_acc}, and Figure~\ref{fig:scatter_steering_wheel} in the Appendix), but they do not allow for a comprehensive description of the driving situation. For example, the effect of vehicle speed may vary not only based on the level of driving automation, but also on the type of road and traffic situation. Therefore, including additional features may not only improve the description of the driving situation but also make the existing features more meaningful by considering their interaction effects.

\fix{Furthermore, it is important to put the results into context and to elaborate on the practical implications this might have. As demonstrated, the approach provides valuable insights into how design artifacts and environmental factors affect drivers' visual attention allocation. The predictions and explanations can guide designers to create interfaces that are less distracting and safer to use. However, even though our approach is superior to related approaches, it is not yet accurate enough to make pixel-precise predictions or to differentiate between minor changes in the driving environment (e.g., driving at 72\,km/h vs. 75\,km/h). To reliably evaluate the effect of such slight changes or to even act as a basis for driver distraction guidelines, the accuracy needs to be increased. Furthermore, we do not consider environmental factors like lightning conditions or street type (e.g., rural road or highway) or UI artifacts like element color and size that might also influence visual attention allocation. Including such features would provide a more holistic picture and probably more accurate predictions. Moreover, drivers tend to self-regulate their willingness to engage in secondary tasks based on the driving task demands \citep{ebel.2022, oviedo-trespalacios.2018b, hancox.2013}. As a result, some interactions occur less frequently in certain driving situations, leading to fewer training data. Therefore, it is likely that prediction accuracy varies across driving situations.}

The presented explanations do not imply causality, and therefore do not represent a complete assessment of drivers' visual attention allocation while being engaged in a secondary touchscreen task. However, the explanations help designers to identify the most informative relationships between input features and model outputs, which assist them in understanding the visual demand predicted by the machine learning model.

\fix{Having shown that this method delivers promising results, the main goal of future iterations is to improve prediction accuracy. First, a more holistic description of the driving situation by providing additional features like lighting conditions, the proximity of surrounding road users, or map data might lead to significant improvements. Second, considering user demographics like age or driving experience might also lead to better accuracy. Finally, a larger dataset is not only likely to benefit the algorithms presented in this work, but would also enable more sophisticated approaches like recurrent neural networks that can capture sequential information embedded in the interaction sequences.}

\section{Conclusion}

In this paper, we propose a machine learning approach that predicts the visual demand of secondary touchscreen interactions while driving, according to the type of interactions that are performed and the associated driving parameters. Our approach generates local and global explanations providing insights how design artifacts and driving parameters affect drivers' visual attention allocation. We evaluate the approach on a real-world driving dataset consisting of 12,142 secondary task engagements. 
Our best model identifies secondary task engagements during which drivers perform a long glance with 68\,\% accuracy and predicts the \ac{TGD} with a mean deviation of 2.4\,s. The analysis of the generated explanations reveals clear differences between the visual demand of specific touchscreen interactions and shows that drivers' visual attention allocation depends on the driving situation. In line with related research~\citep{Risteska.2021, Tivesten.2014}, we show that drivers modulate their visual attention allocation based on the vehicle speed and the level of driving automation.

Our key contributions address many points that previous approaches \citep{Risteska.2021, Large.2017, Janssen.2015, Victor.2014} have identified as desirable: (1) The approach leverages continuously collected large-scale real-world data providing realistic predictions of drivers' visual attention allocation during secondary task engagements. (2) The approach can easily be adjusted to incorporate additional features and to predict various metrics in addition to TGD and long glance probability (e.g., number of glances, total eyes off-road time, mean glance duration). (3) The local and global explanations provide detailed insight into the impact design artifacts and scenario parameters have on driver distraction prediction. (4) The approach can inform designers about potential implications their design may have and can guide them to design in-vehicle touchscreen interfaces that are safe to use.

\bibliography{VisualDemandPrediction.bib}
\bibliographystyle{cas-model2-names}

\appendix
\section{Appendix}\label{ch:Appendix}
\subsection{Definitions}\label{ch:AppendixDefinitions}
\textbf{Definition 1:} An \textit{interaction sequence} $I = (i_n)_{n=1}^{N}$ is a sequence of touchscreen interactions recorded during one trip, where $i_n$ is a single touchscreen interaction performed by a user and $N$ denotes the number of interactions of $I$. A touchscreen interaction $i = (t, e, p, c)$ is composed of its timestamp $t$, element type $e$, gesture type $p$ and coordinate pair $c = (x,y)$. Within $I$, the duration between two successive interactions $t(i_{n+1}) - t(i_n)$ must be smaller than $\Delta t_{max}$ such that $t(i_{n+1}) - t(i_n) \leq \Delta t_{max}$.

\textbf{Definition 2:} A \textit{glance sequence} $G = (g_n)_{n=1}^N$ is a sequence of non-overlapping intervals of driver glances, where $g_n$ is a single glance performed by a user and $N$ denotes the number of glances of $G$. Each glance $g_n = (t^{s}, t^{e}, r)_n$ is composed of its start time $t^s$, end time $t^e$, and AOI $r$, describing where looked at between $t^s$ and $t^e$. For all glances of a but the first of a trip, the start time is equal to the end time of the preceding glance $t_s(g_{n}) = t_s(g_{n-1})$.

\textbf{Definition 3:} A \textit{driving sequence} $D=(d_n)_{n=1}^N$ is a sequence of driving data observations, where $d_n$ is a single observation and $N$ denotes the number of observations of $D$. Each observation is defined as $d_n = (t, v, \theta, a_{ACC},a_{SA})_n$, where $t$ represents the timestamp, $v$ the vehicle speed, $\theta$ the steering wheel angle, $a_{ACC}$ and $a_{SA}$ the status of the ACC and SA respectively.

\textbf{Definition 4:} A \textit{secondary task engagement} $S$ is defined as an interaction sequence and its corresponding glance sequence and driving sequence $S = (I,G,D)$. We consider all driving observations starting before the first interaction until after the last interaction such that $t(i_{1}) - t_b < t(d_n) < t(i_{N}) + t_b$. Where $t_b$ represents a buffer duration. Regarding the glance sequence $G$, we consider all glances whose start time or end time falls in between the first and last interaction of $I$ such that $t(i_{1}) < t^s(g_n) < t(i_{N}) \vee t(i_{1}) < t^e(g_n) < t(i_{N})$.

\textbf{Problem 1:} We define the \textit{long glance prediction task} as the problem of identifying all secondary task engagements $S$ in which a long glance occurs such that for any $g_n \in G $, $\Delta t_g = t^e(g_n) - t^s(g_n) > 2\,s$ given the according interaction sequence $s^{i}$ and driving sequence $s^{d}$

\textbf{Problem 2:} We define the \textit{total glance duration prediction task} as the problem of predicting the TGD toward the center stack touchscreen during a secondary task engagement $S$.

\subsection{Hyperparameter Optimization}\label{ch:AppendixHyperOpt}
In the following we report the results of the hyperparameter optimization for each of the individual models.

\subsubsection{Random Forest Models}
The Implementation and the descriptions based on the \href{https://scikit-learn.org/stable/index.html}{scikit-learn} python package.

    \textbf{n\_estimators} = [100, 200, 400, 800, 1200, 1600, 2000] -- The number of trees in the forest.
    
    \textbf{max\_features} = ['auto', 'sqrt'] -- Number of features to consider when looking for the best split
    
    \textbf{max\_depth} = [10, 20, 30, 40, 60, 80, 100] -- Maximum depth of the tree. 
    
    \textbf{min\_samples\_split} = [2, 5, 10] -- Minimum number of samples required to split an internal node.
    
    \textbf{min\_samples\_leaf} = [1, 2, 4] -- Minimum number of samples required to be at a leaf node.
    
    \textbf{bootstrap} = [True, False] -- Whether bootstrap samples are used when building trees. \\

\begin{table}[pos=htbp!]
	\caption{Sets of best performing parameters for the Random Forest models.}
	\label{tab:resultRF}
	\centering
	\begin{tabular}{lrr} 
		\toprule
		Feature		& Long Glance Prediction & TGD Prediction\\
		\toprule
		n\_estimators & 200 & 1600 \\
		max\_features   & auto & auto \\
		max\_depth & 10 & 60 \\
		min\_samples\_split & 5 & 2 \\
		min\_samples\_leaf   & 2 & 4 \\
		bootstrap  & True & True \\ 
		\bottomrule
	\end{tabular}
\end{table}

\subsubsection{XGBoost Models}
The Implementation and the descriptions are based on the
\href{https://xgboost.readthedocs.io/en/stable/python/python\_api.html}{XGBoost} python package.

\textbf{n\_estimators} = [20, 100, 500, 1000, 5000, 10000, 20000] -- Number of boosting rounds.

\textbf{subsample} = [0.2,0.4,0.6,0.8,1] -- Subsample ratio of the training instance.

\textbf{max\_depth} = [5,10,50,100] -- Maximum tree depth for base learners.

\textbf{learning\_rate} = [0.0005, 0.001, 0.01, 0.1, 1] -- Boosting learning rate (xgb’s “eta”)

\textbf{colsample\_bytree} = [0.2,0.4,0.6,0.8,1] -- Subsample ratio of columns when constructing each tree.

\textbf{colsample\_bylevel} = [0.2,0.4,0.6,0.8,1] -- Subsample ratio of columns for each level.

\begin{table}[pos=htbp!]
\small
	\caption{Sets of best performing parameters for the XGBoost models.}
	\label{tab:resultXGBoost}
	\centering
	\begin{tabular}{lrr} 
		\toprule
		Feature		& Long Glance Prediction & TGD Prediction\\
		\toprule
		n\_estimators & 5000 & 5000 \\
		subsample   & 0.6 & 0.8 \\
		max\_depth & 10 & 10 \\
		min\_child\_weight & 4 & 10 \\
		learning\_rate & 0.01 & 0.0005 \\
		colsample\_bytree   & 0.2 & 0.6 \\
		colsample\_bylevel  & 0.2 & 1 \\ 
		\bottomrule
	\end{tabular}
\end{table}

\subsubsection{Feedforward Neural Networks}

The Implementation and the hyperparameter optimization was performed using the \href{https://keras.io}{Keras} API. It needs to be noted that the different hyperparameter combinations did not show large differences in their predictive performance.

\textbf{n\_hidden\_layers} = [1,2,3,4,5] -- Number of layers between the input and output layer of the neural network

\textbf{n\_neurons} = [32, 64, 128, 256, 512]) -- Number of neurons per layer.

\textbf{activation} = ["relu", "sigmoid"] -- The activation function of the neurons in the respective layer.

\textbf{drop\_out} = [0,0.1,0.2,0.3] -- The probability at which random units are set to zero during training.

\textbf{learning\_rate} = [0.01,0.001,0.0001] -- Initial learning rate of the ADAM optimizer.

\begin{table}[pos=htbp!]
\small
	\caption{Sets of best performing parameters for the FNN models.}
	\label{tab:resultFF}
	\centering
	\begin{tabular}{lrr} 
		\toprule
		Feature		& Long Glance Prediction & TGD Prediction\\
		\toprule
		n\_hidden\_layers           & 3         & 1 \\
		learning\_rate              & 0.0001     & 0.001 \\
        n\_neurons layer 1          & 512        & 512 \\
		activation layer 1          & sigmoid   & relu \\
		drop\_out layer 1           & 0.3       & 0.1 \\
		n\_neurons layer 2          & 64       & - \\
		activation layer 2          & relu   & - \\
		drop\_out layer 2           & 0.1       & - \\
		n\_neurons layer 3          & 256       & - \\
		activation layer 3          & sigmoid   & - \\
		drop\_out layer 3           & 0.1       & - \\
		\bottomrule
	\end{tabular}
\end{table}

\newpage
\subsection{Dataset Summary Statistics}\label{ch:AppendixSummaryStats}
Table \ref{tab:SummaryStats} provides an overview of all features.

\begin{table}[pos=htpb!] \centering
\footnotesize
  \caption{Dataset Summary Statistics} 
  \label{tab:SummaryStats} 
\begin{tabular}{@{\extracolsep{5pt}}lrrrrrrr} 
\\[-1.8ex]\hline 
\hline \\[-1.8ex] 
Statistic & \multicolumn{1}{c}{Mean} & \multicolumn{1}{c}{St. Dev.} & \multicolumn{1}{c}{Min} & \multicolumn{1}{c}{$Q_1$(25)} & \multicolumn{1}{c}{Median} & \multicolumn{1}{c}{$Q_3$(75)} & \multicolumn{1}{c}{Max} \\ 
\hline \\[-1.8ex] 
Number of interactions & 4.431 & 4.993 & 1 & 1 & 3 & 5 & 41 \\ 
Number of tap gestures & 3.814 & 4.495 & 0 & 1 & 2 & 5 & 40 \\ 
Number of drag gestures & 0.363 & 1.324 & 0 & 0 & 0 & 0 & 31 \\ 
Number of multitouch gestures & 0.240 & 1.108 & 0 & 0 & 0 & 0 & 26 \\ 
Average glance duration in ms& 1,441.491 & 929.736 & 120.000 & 960.000 & 1,241.000 & 1,659.000 & 26,801.000 \\ 
Number of glances & 4.367 & 4.998 & 1 & 1 & 3 & 6 & 50 \\ 
number of long glances & 0.569 & 1.102 & 0 & 0 & 0 & 1 & 13 \\ 
Total glance duration in ms & 5,742.751 & 7,049.487 & 120.000 & 1,590.500 & 3,499.000 & 7,354.000 & 262,416.000 \\ 
average speed in km/h & 70.516 & 36.935 & 0.633 & 40.881 & 66.230 & 96.567 & 209.883 \\ 
ACC active & 0.206 & 0.404 & 0 & 0 & 0 & 0 & 1 \\ 
SA active & 0.099 & 0.299 & 0 & 0 & 0 & 0 & 1 \\ 
AppIcon interactions& 0.196 & 0.586 & 0 & 0 & 0 & 0 & 13 \\ 
CoverFlow interactions& 0.038 & 0.434 & 0 & 0 & 0 & 0 & 16 \\ 
Unknown interactions& 0.049 & 0.450 & 0 & 0 & 0 & 0 & 23 \\ 
Other interactions& 0.731 & 1.568 & 0 & 0 & 0 & 1 & 37 \\ 
List interactions& 0.518 & 1.652 & 0 & 0 & 0 & 0 & 31 \\ 
Tab interactions& 0.385 & 1.335 & 0 & 0 & 0 & 0 & 35 \\ 
ControlBar interactions& 0.012 & 0.135 & 0 & 0 & 0 & 0 & 4 \\ 
Button interactions& 0.640 & 1.515 & 0 & 0 & 0 & 1 & 36 \\ 
Homebar interactions& 0.892 & 2.032 & 0 & 0 & 0 & 1 & 36 \\ 
Slider interactions & 0.015 & 0.228 & 0 & 0 & 0 & 0 & 9 \\ 
ClickGuard interactions& 0.058 & 0.313 & 0 & 0 & 0 & 0 & 8 \\ 
PopUp interactions& 0.030 & 0.232 & 0 & 0 & 0 & 0 & 9 \\ 
Keyboard interactions& 0.184 & 1.435 & 0 & 0 & 0 & 0 & 31 \\ 
Map interactions& 0.508 & 2.181 & 0 & 0 & 0 & 0 & 39 \\ 
RemoteUI interactions& 0.173 & 1.179 & 0 & 0 & 0 & 0 & 31 \\ 
Browser interactions & 0.002 & 0.093 & 0 & 0 & 0 & 0 & 8 \\ 
\hline \\[-1.8ex] 
\end{tabular} 
\end{table} 

\newpage
\subsection{Steering Wheel Feature Dependence Plot}

	\begin{figure}[pos=htpb!]
		\centering
		\subfloat[Long Glance Prediction Sample. ]{\includegraphics[width=0.49\linewidth]{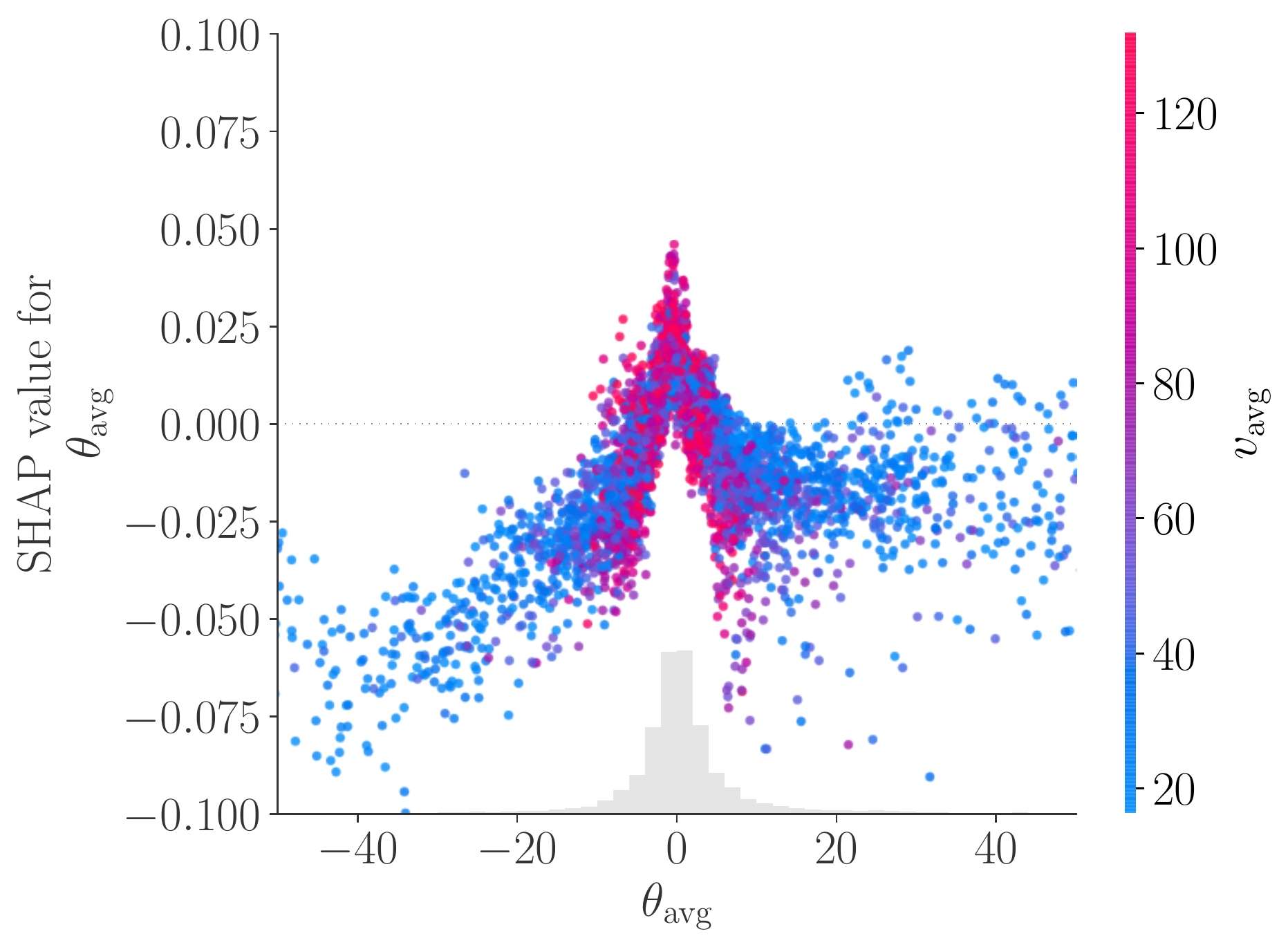}\label{fig:scatter_plot_classification_steering}}
		\subfloat[Total Glance Duration Prediction Sample.]{\includegraphics[width=0.49\linewidth]{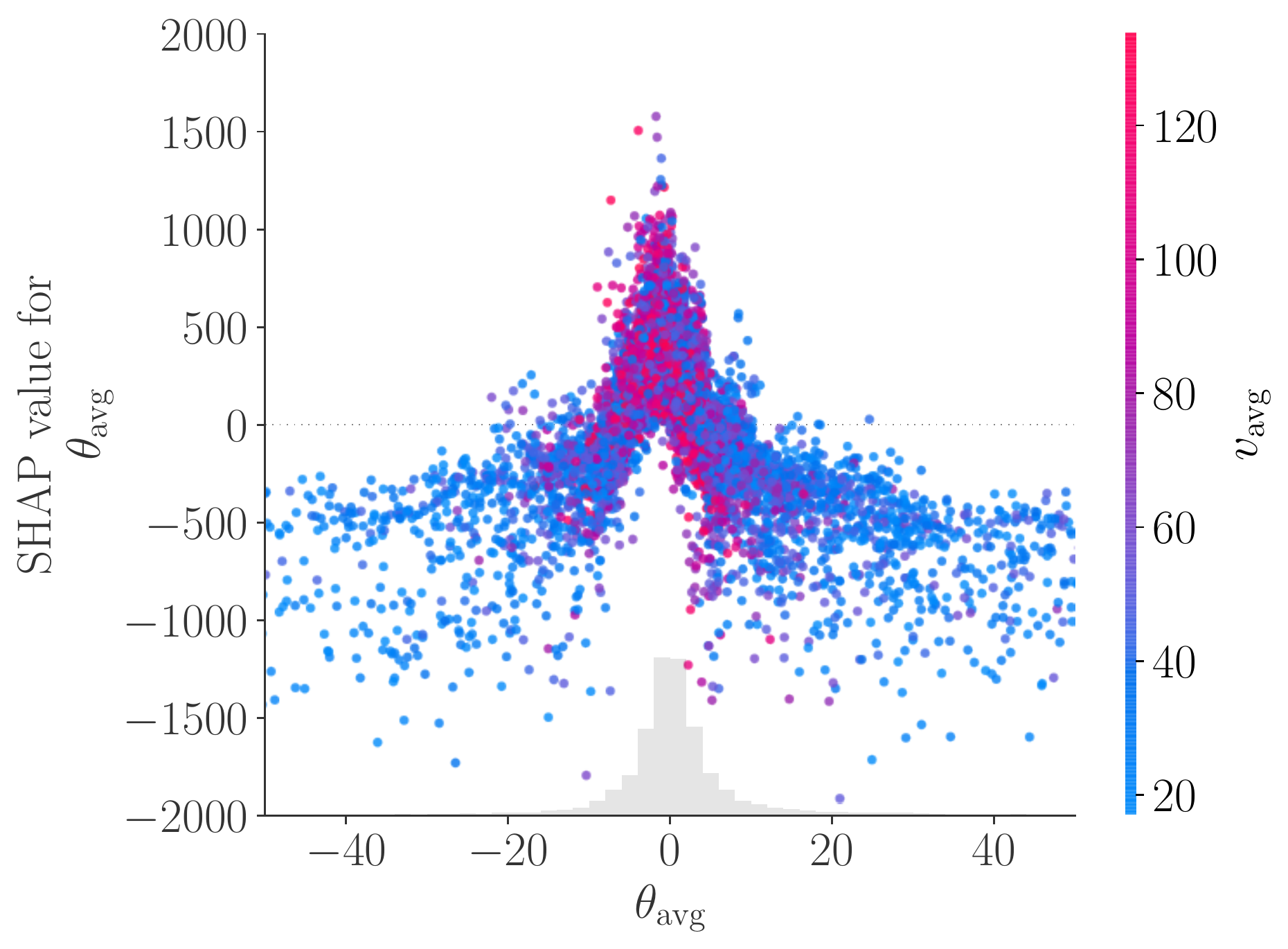}\label{fig:scatter_plot_regression_steering}}
		\caption{Feature dependence plots of the steering wheel angle and its interaction with the vehicle speed for the long glance classification and TGD model. }
		\label{fig:scatter_steering_wheel} 
	\end{figure}

\end{document}